\documentclass[12pt]{article}

\usepackage{epsfig,latexsym,amsfonts,amsmath,amsthm,amssymb,amsbsy,multirow,slashed,wasysym,textcomp,subfigure,hyperref,wrapfig,datetime,comment,mathtools,cancel,cite,tensor}
\usepackage{mathrsfs}
\usepackage[font={footnotesize,sl},bf]{caption}

\usepackage{enumitem}

\usepackage{pict2e}
\makeatletter
\newcommand{\loplus}{\mathbin{\mathpalette\dog@lsemi{+}}}
\newcommand{\dog@lsemi}[2]{\dog@semi{#1}{#2}{270,90}}
\newcommand{\dog@semi}[3]{%
  \begingroup
  \sbox\z@{$\m@th#1#2$}%
  \setlength{\unitlength}{\dimexpr\ht\z@+\dp\z@\relax}%
  \makebox[\wd\z@]{\raisebox{-\dp\z@}{%
    \begin{picture}(1,1)
    \linethickness{\variable@rule{#1}}
    \roundcap
    \put(0.5,0.5){\makebox(0,0){\raisebox{\dp\z@}{$\m@th#1#2$}}}
    \put(0.5,0.5){\arc[#3]{0.5}}
    \end{picture}%
  }}%
  \endgroup
}
\newcommand{\variable@rule}[1]{%
  \fontdimen8  
  \ifx#1\displaystyle\textfont3\else
    \ifx#1\textstyle\textfont3\else
      \ifx#1\scriptstyle\scriptfont3\else
        \scriptscriptfont3\relax
  \fi\fi\fi
}
\makeatother
\def\ee{\end{equation}}
\def\bea{\begin{eqnarray}}
\def\eea{\end{eqnarray}}
\newcommand{\beq}{\begin{eqnarray}}
\newcommand{\eqq}{\end{eqnarray}}
 \newcommand{\badat}{\begin{alignedat}}
 \newcommand{\eadat}{\end{alignedat}}
 
\newcommand{\eal}[1]{\be \begin{aligned} #1 \end{aligned}\end{equation}} 
\newcommand{\eqn}[1]{\be #1 \end{equation}} 
\newcommand{\eqa}[1]{\bea  #1\end{eqnarray}}

\long\def\new#1\endnew{{\bf #1}}		
\long\def\del#1\enddel{}

\def\del{\partial}

\def\12{\frac{1}{2}}
\def\32{\frac{3}{2}}
\usepackage{xcolor}

\newcommand{\p}{\partial}

\def\bz{{\bar z}}
\def\bw{{\bar w}}

\numberwithin{equation}{section} % equation numbers follow sections

\begin{document}
\begin{titlepage}
  \thispagestyle{empty}

  \begin{flushright}
  \end{flushright}

\vskip3cm

  \begin{center}  
{\LARGE\textbf{BMS Flux Algebra in Celestial Holography}}
 % \vskip1cm \today, \currenttime

\vskip1cm

   \centerline{Laura Donnay\footnote{laura.donnay@tuwien.ac.at}, Romain Ruzziconi\footnote{romain.ruzziconi@tuwien.ac.at}}

\vskip1cm

{\it{Institute for Theoretical Physics, TU Wien}}\\
{{\it Wiedner Hauptstrasse 8–10/136, A-1040 Vienna, Austria}}

\end{center}

\vskip1cm

\begin{abstract}
Starting from gravity in asymptotically flat spacetime, the BMS momentum fluxes are constructed. These are non-local expressions of the solution space living on the celestial Riemann surface. They transform in the coadjoint representation of the extended BMS group and correspond to Virasoro primaries under the action of bulk superrotations. The relation between the BMS momentum fluxes and celestial CFT operators is then established: the supermomentum flux is related to the supertranslation operator and the super angular momentum flux is linked to the stress-energy tensor of the celestial CFT. The transformation under the action of asymptotic symmetries and the OPEs of the celestial CFT currents are deduced from the BMS flux algebra.
\end{abstract}

\end{titlepage}

\tableofcontents

\section{Introduction}

Celestial holography aims at establishing a holographic  description of quantum gravity in four-dimensional asymptotically flat spacetime in terms of a two-dimensional conformal field theory, called celestial CFT (or CCFT for short), living on the boundary celestial Riemann surface. This program exploits the richness of the asymptotic symmetry structure of the spacetime \cite{Bondi:1962px,Sachs:1962wk,Sachs:1962zza,deBoer:2003vf,Barnich:2010eb,Strominger:2013jfa} to constrain the potential candidate for the celestial dual theory.
 Conformal symmetries of the CCFT are induced by superrotations which are part of the (extended) Bondi-Metzner-Sachs (BMS) asymptotic symmetries in the bulk theory \cite{Barnich:2010eb,Barnich:2009se,Barnich:2011ct,Kapec:2014opa, Fotopoulos:2019vac}. In celestial holography, each scattering particle in the bulk spacetime is associated to an operator that lives on the boundary celestial Riemann surface. In those terms, soft theorems in the bulk spacetime, corresponding to Ward identities of the gravitational $S$-matrix for the extended BMS symmetries, are implemented by $2d$ currents in the CCFT; see \cite{Strominger:2017zoo} for a review. In particular, the supertranslation current has Ward identities that are equivalent to the leading soft graviton theorem \cite{Strominger:2013jfa,He:2014laa}, while the sub-leading soft theorem is obtained by the insertion of a holographic stress-tensor \cite{Kapec:2016jld}. A natural basis to describe massless asymptotic particles in celestial holography can be obtained  by applying a Mellin transform with respect to the energy of the external particle, which maps energy eigenstates to boost eigenstates and hence makes the conformal properties more manifest \cite{deBoer:2003vf,Pasterski:2016qvg,Pasterski:2017ylz,Pasterski:2017kqt,Cheung:2016iub}.

In \cite{Barnich:2021dta}, the coadjoint representation of the BMS group in four dimensions has been constructed. It acts on a set of conformal fields that have been identified with local expressions of the solution space of non-radiative asymptotically flat spacetimes at null infinity through a (pre-)momentum map. In presence of radiation, the transformation of the gravitational solution space becomes more complicated and the coadjoint representation of the BMS group is not sufficient to describe it. Furthermore, the BMS surface charges become non-integrable \cite{Wald:1999wa} and one needs additional inputs to select a meaningful integrable part. The algebra requires the use the modified Barnich-Troessaert bracket \cite{Barnich:2011mi}, leading to a field-dependent $2$-cocycle (see also \cite{Distler:2018rwu} for a detailed analysis from double-soft limits of amplitudes). As shown in \cite{Freidel:2021fxf, Freidel:2021cbc}, the latter can be re-absorbed in the definition of the modified bracket by using the Noetherian split between integrable and non-integrable parts.

Instead of working with local expressions of the solution space at finite value of retarded time $u$, one could consider fluxes that correspond to integrated expressions over $u$. This point of view is closer to the spirit of celestial holography where the retarded time does not appear explicitly in the CCFT. It was shown in \cite{Campiglia:2020qvc, Compere:2020lrt} that, provided one chooses an appropriate integrable part in the BMS surface charges, the algebra of associated BMS fluxes closes under the standard bracket. The prescription that we consider here is based on \cite{Compere:2020lrt,Compere:2018ylh} and has the following important properties: (i) the flux algebra closes under the standard bracket, (ii) the fluxes vanish when evaluated on vacuum solutions (namely solutions with identically vanishing Riemann tensor that are constructed in \cite{Compere:2016jwb , Compere:2016hzt , Compere:2016gwf}), (iii) the fluxes vanish for non-radiative spacetime solutions, (iv) the fluxes are finite provided one chooses the appropriate falloffs in $u$.

In this paper, we identify some non-local combinations of the solution space of four-dimensional asymptotically flat spacetimes that transform in the coadjoint representation of the extended BMS group in presence of radiation. We call these expressions the BMS momentum fluxes since they are involved in the BMS fluxes discussed in \cite{Compere:2020lrt}. The inclusion of the superrotations in the analysis requires a meticulous treatment of the $2d$ Liouville stress-tensor discussed in \cite{Compere:2016jwb,Compere:2018ylh}. In a second step, we propose a new prescription to split the fluxes into soft and hard parts, so that the associated soft and hard phase spaces factorize. We then relate the soft BMS momentum fluxes with the supertranslation operator and the stress-tensor of the CCFT. We provide the precise expressions of these CCFT currents in terms of the bulk metric and deduce their transformation laws under extended BMS transformations. Finally, from the BMS flux algebra, we deduce the OPEs of the BMS momentum fluxes and recover the OPEs of the CCFT operators.

\section{Asymptotically flat spacetimes}
\label{Asymptotically flat spacetimes}

In this Section, we describe the bulk side of the celestial holographic description by reviewing the analysis of four-dimensional asymptotically flat spacetimes at null infinity, denoted $\mathscr{I}^+$, in Bondi gauge \cite{Bondi:1962px,Sachs:1962wk,Sachs:1962zza}; see \cite{Herfray:2020rvq,Herfray:2021xyp} for an intrinsic conformally invariant geometrical description of null infinity. We mainly follow the notations and conventions of \cite{Barnich:2010eb,Compere:2018ylh}. In Bondi coordinates $(u, r, x^A)$, $x^A = (z, \bar{z})$, the spacetime metric reads as 
\begin{equation}
    ds^2 = e^{2\beta} \frac{V}{r} {d}u^2 - 2 e^{2\beta}{d}u {d}r + g_{AB} ({d}x^A - U^A {d}u)(dx^B - U^B {d}u),
    \label{Bondi gauge metric}
\end{equation} where $\beta$, $V$, $g_{AB}$, $U^A$ are functions of $(u,r, x^A)$ and the transverse metric $g_{AB}$ satisfies the determinant condition
\begin{equation}
    \partial_r [r^{-4} \det (g_{AB})] = 0.
\end{equation} We consider the asymptotically flat spacetimes satisfying the boundary conditions
\begin{equation}
    \beta = \mathcal{O}(r^{-1}), \quad \frac{V}{r} = -1 + \mathcal{O}(r^{-1}), \quad U^A = \mathcal{O}(r^{-1}), \quad g_{AB} = r^2 \mathring{q}_{AB} + r C_{AB} + \mathcal{O}(r^{-1}) ,
    \label{BC Bondi gauge}
\end{equation} where
\begin{equation}
    \mathring{q}_{AB} dx^A dx^B = 2 (\Omega_S \bar{\Omega}_S)^{-1} dz d\bar{z}, \qquad \Omega_S = \frac{1+ z\bar{z}}{\sqrt{2}} = \bar{\Omega}_S 
    \label{2 sphere metric}
\end{equation} is unit sphere metric and 
$C_{AB}(u,x)$ is a $2$-dimensional symmetric traceless tensor called the asymptotic shear. Let us make some comments on the choice of falloffs \eqref{BC Bondi gauge}, \eqref{2 sphere metric}:
\begin{itemize}[leftmargin=*]
    \item We allow for possible puncture singular violations of the above boundary conditions to accommodate with the $\text{Witt} \oplus \overline{\text{Witt}}$ superrotations symmetries \cite{Barnich:2010eb,Barnich:2009se,Barnich:2011ct} that we discuss below. In particular, we consider the topology of the $2$-punctured sphere as celestial Riemann surface $\mathcal{S} \simeq \mathscr{I}^+ / \mathbb{R}$ \cite{Barnich:2021dta}. 
    \item Possible relaxations of the above boundary conditions have been considered recently in the literature allowing for variations of the transverse boundary metric. These lead to enhancement of the asymptotic group with Diff($S^2$) superrotations \cite{Campiglia:2015yka,Campiglia:2014yka,Compere:2018ylh,Flanagan:2019vbl} and/or Weyl rescaling symmetries \cite{Barnich:2010eb, Barnich:2016lyg , Barnich:2019vzx , Freidel:2021fxf} (see also \cite{Ruzziconi:2020cjt} for a review). While the case that we discuss here is the most natural to study the celestial holography since it readily implies the conformal symmetries on the celestial Riemann surface, we will comment on these extensions in the discussion section.  
    
    \item In the above conditions, we set the order $r^0$ in the expansion of $g_{AB}$ to zero. Turning on this term would bring some $\log r$ terms in the expansion that we want to avoid \cite{Tamburino:1966zz, Winicour1985LogarithmicAF,Barnich:2010eb}. For discussions on polyhomogeneous spacetimes, see e.g. \cite{Andersson:1993we, Chrusciel:1993hx , Andersson:1994ng, Ashtekar:1996cm , Friedrich:2017cjg , Angelopoulos:2017iop, Kroon:1998tu, Godazgar:2020peu}.  
\end{itemize}
Solving Einstein's equations in vaccuum with vanishing cosmological constant for the boundary conditions \eqref{BC Bondi gauge} yields the following expansions \cite{Tamburino:1966zz,Barnich:2010eb}: 
\begin{equation}
\begin{split}
\frac{V}{r} &= -1 + \frac{2M}{r} + \mathcal{O}(r^{-2}) ,\qquad 
\beta = \frac{1}{r^2} \left[ - \frac{1}{32}C^{AB} C_{AB} \right] +  \mathcal{O}(r^{-3}), \\
g_{AB} &= r^2 \mathring{q}_{AB} + r C_{AB} + \mathcal{O}(r^{-1}) ,\\
U^A &= -\frac{1}{2 r^2} D_B C^{AB} -\frac{2}{3} \frac{1}{r^3} \left[ N^A - \frac{1}{2} C^{AB} D^C C_{BC} \right] + \mathcal{O}(r^{-4}),
\end{split}
\label{fall-off}
\end{equation} 
where $M = M(u,x)$ is the Bondi mass aspect, $N_A = N_A (u,x)$ is the angular momentum aspect. The 2-sphere indices in \eqref{fall-off} are lowered an raised with $\mathring{q}_{AB}$ and its inverse, and $D_A$ is the Levi-Civita connection on the celestial Riemann surface associated to $\mathring{q}_{AB}$. The Bondi mass and angular momentum aspects satisfy the time evolution equations
\begin{equation}
\begin{split}
\partial_u M &= - \frac{1}{8} N_{AB} N^{AB} + \frac{1}{4} D_A D_B N^{AB} , \\ 
\partial_u N_A &= D_A M + \frac{1}{16} D_A (N_{BC} C^{BC}) - \frac{1}{4} N^{BC} D_A C_{BC}  -\frac{1}{4} D_B (C^{BC} N_{AC} - N^{BC} C_{AC}) \\
&\quad - \frac{1}{4} D_B D^B D^C C_{AC}+ \frac{1}{4} D_B D_A D_C C^{BC} ,
\label{EOM1} 
\end{split}
\end{equation} 
with $N_{AB} = \partial_u C_{AB}$ the Bondi news tensor.

The residual diffeomorphisms that preserve the Bondi gauge \eqref{Bondi gauge metric} and falloff conditions \eqref{BC Bondi gauge} are generated by vectors fields $\xi = \xi^u \partial_u + \xi^z \partial + \xi^{\bar{z}} \bar{\partial}+ \xi^r \partial_r$ whose components read as
\begin{equation}
\begin{split}
    &\xi^u =(\Omega_S\bar{\Omega}_S)^{-\frac{1}{2}} \mathcal{T} + \frac{u}{2} (D_z \mathcal{Y} +  D_{\bar{z}} \bar{\mathcal{Y}}) , \\
    &\xi^z = \mathcal{Y} + \mathcal{O}(r^{-1}) \quad \xi^{\bar{z}} = \bar{\mathcal{Y}} + \mathcal{O}(r^{-1}), \\ 
    &\xi^r = - \frac{r}{2} (D_z \mathcal{Y} +  D_{\bar{z}} \bar{\mathcal{Y}}) + \mathcal{O}(r^0),
    \end{split}
    \label{AKV Bondi}
\end{equation} where $\mathcal{T}= \mathcal{T}(z, \bar{z})$ is the supertranslation parameter and $\mathcal{Y} = \mathcal{Y}(z)$, $\bar{\mathcal{Y}} = \bar{\mathcal{Y}}(\bar{z})$ are the superrotation parameters satisfying the conformal Killing equation
\begin{equation}
    D_{\bar{z}} \mathcal{Y} = 0, \qquad D_z \bar{\mathcal{Y}} = 0.
\end{equation} Using the modified Lie bracket $[\xi_1, \xi_2]_\star =[\xi_1, \xi_2] - \delta_{\xi_1}\xi_2 + \delta_{\xi_2} \xi_1$ where the last two terms take into account the field-dependence of the asymptotic Killing vectors \eqref{AKV Bondi} at subleading order in $r$ \cite{Schwimmer:2008yh , Barnich:2010eb}, the asymptotic Killing vectors \eqref{AKV Bondi} satisfy the commutation relations
\begin{equation}
    [\xi (\mathcal{T}_1, \mathcal{Y}_1, \bar{\mathcal{Y}}_1), \xi (\mathcal{T}_2, \mathcal{Y}_2, \bar{\mathcal{Y}}_2)]_\star = \xi (\mathcal{T}_{12}, \mathcal{Y}_{12}, \bar{\mathcal{Y}}_{12}),
    \label{commutation relations 1}
\end{equation} with
\begin{equation}
    \mathcal{T}_{12} = \mathcal{Y}_1 \partial \mathcal{T}_2 - \frac{1}{2} \partial \mathcal{Y}_1 \mathcal{T}_2 - (1 \leftrightarrow 2) + c.c., \quad \mathcal{Y}_{12} = \mathcal{Y}_1 \partial \mathcal{Y}_2 - (1 \leftrightarrow 2), \quad \bar{\mathcal{Y}}_{12} = \bar{\mathcal{Y}}_1 \bar{\partial} \bar{\mathcal{Y}}_2 - (1 \leftrightarrow 2)
    \label{commutation relations 2}
\end{equation} where $c.c.$ stands for complex conjugate terms. This corresponds to the extended BMS algebra, namely
\begin{equation}
   \mathfrak{bms}_4 = (\text{Witt} \oplus \overline{\text{Witt}}) \loplus \mathfrak{s}^*,
\end{equation} where $\mathfrak{s}^*$ stands for the (possibly singular) supertranslations.  

For convenience, we introduce the notations $f = (\Omega_S\bar{\Omega}_S)^{-\frac{1}{2}}\mathcal{T} + \frac{u}{2} (D_z \mathcal{Y} +  D_{\bar{z}} \bar{\mathcal{Y}})$ and $Y^A = (\mathcal{Y}, \bar{\mathcal{Y}})$. Under residual gauge diffeomorphisms \eqref{AKV Bondi}, the solution space transforms infinitesimally as
\begin{equation}
    \begin{split}
\delta_{(f,Y)} C_{AB} &= [f \partial_u + \mathcal{L}_Y - \frac{1}{2} D_C Y^C ] C_{AB} - 2 D_A D_B f + \mathring{q}_{AB} D_C D^C f,\\
\delta_{(f,Y)} N_{AB} &= [f\partial_u + \mathcal{L}_Y] N_{AB} - (D_A D_B D_C Y^C - \frac{1}{2} \mathring{q}_{AB} D_C D^C D_D Y^D),\\
\delta_{(f,Y)} M &= [f \partial_u + \mathcal{L}_Y + \frac{3}{2} D_C Y^C] M \\
& + \frac{1}{8} D_C D_B D_A Y^A C^{BC} + \frac{1}{4} N^{AB} D_A D_B f + \frac{1}{2} D_A f D_B N^{AB},\\
\delta_{(f,Y)} N_A &= [f\partial_u + \mathcal{L}_Y + D_C Y^C] N_A + 3 M D_A f - \frac{3}{16} D_A f N_{BC} C^{BC}  \\
&\quad - \frac{1}{32} D_A D_B Y^B C_{CD}C^{CD} + \frac{1}{4} (2 D^B f + D^B D_C D^C f) C_{AB}  \\
&\quad - \frac{3}{4} D_B f (D^B D^C C_{AC} - D_A D_C C^{BC}) + \frac{3}{8} D_A (D_C D_B f C^{BC}) \\
&\quad + \frac{1}{2} (D_A D_B f - \frac{1}{2} D_C D^C f \mathring{q}_{AB}) D_C C^{BC} + \frac{1}{2} D_B f N^{BC} C_{AC}.   
\end{split}
\label{transformation on the solution space}
\end{equation}

As one can see from the second expression above, the Bondi news $N_{AB}$ transforms inhomogeneously under superrotations. As discussed in \cite{Compere:2018ylh}, one can define the physical news $\hat{N}_{AB}$ as \begin{equation}
   \hat{N}_{AB}(u,x) = N_{AB}(u,x) - N_{AB}^{vac}(x) ,
   \label{physical}
\end{equation} with
\begin{equation}
    N_{AB}^{vac}(x) =\left[ \frac{1}{2}D_A\Phi D_B\Phi - D_A D_B \Phi \right]^{TF} , \qquad \Phi(z, \bar{z}) = \varphi (z) + \bar{\varphi}(\bz) +  \ln ( \Omega_S \bar{\Omega}_S) ,
    \label{Liouville stress tensor}
\end{equation}
where $TF$ stands for the trace-free part. $N_{AB}^{vac}$ is the trace-free part of the stress-tensor for a $2d$ Euclidean Liouville theory living on the celestial Riemann surface with Lagrangian 
\begin{equation}
    L[\Phi] = \sqrt{\mathring{q}} \left( \frac{1}{2} D_A \Phi D^A \Phi + \mathring{R} \Phi \right).
\end{equation} The Liouville scalar field $\Phi$ is called the ``superboost field'' and it encodes the refraction/velocity kick memory effects \cite{Compere:2018ylh}. It satisfies the equation of motion
\begin{equation}
    \Box \Phi = \mathring{
    R} = 2 \Omega_S\bar{\Omega}_S \partial \bar{\partial} \ln (\Omega_S\bar{\Omega}_S ) = 2
    \label{Liouville equation}
\end{equation} and transforms as 
\begin{equation}
     \delta_{(f,Y)} \Phi = Y^A D_A \Phi + D_A Y^A 
     \label{transfo Liouville}
\end{equation} or, equivalently, $\delta_{(f,Y)} \varphi = \mathcal{Y} \partial \varphi + \partial \mathcal{Y}, \, \delta_{(f,Y)} \bar{\varphi} = \bar{\mathcal{Y}} \bar{\partial} \bar{\varphi} + \bar{\partial} \bar{\mathcal{Y}}$, under residual gauge diffeomorphisms \eqref{AKV Bondi}. Notice that the Liouville equation \eqref{Liouville equation} is consistent with the action of the symmetries since $\delta_{f,Y} (\Box \Phi - \mathring{
    R}) = (\mathcal{L}_Y+ D_A Y^A) (\Box \Phi - \mathring{
    R})$. As a consequence of \eqref{Liouville equation} and \eqref{transfo Liouville}, the Liouville stress-tensor \eqref{Liouville stress tensor} satisfies $D^A N_{AB}^{vac} = 0$
and transforms as 
\begin{equation}
\delta_{(f,Y)}N^{vac}_{AB}  = \mathcal{L}_Y N_{AB}^{vac} - (D_A D_B D_C Y^C)^{TF}.    
\label{transfo of the 2d Liouville tens}
\end{equation}
One can show that it is related to the trace-free part of the Geroch tensor $\rho_{AB}$   \cite{Geroch1977,Ashtekar:2014zsa,Campiglia:2020qvc,Nguyen:2020hot}. The interest of the physical news \eqref{physical} is that it transforms homogeneously, i.e.
\begin{equation}
    \delta_{(f,Y)} \hat{N}_{AB} = [f\partial_u + \mathcal{L}_Y] \hat{N}_{AB},
\end{equation} so that $\hat{N}_{AB} = 0$ is a meaningful condition to impose in presence of superrotations to define non-radiative spacetimes.

In addition to the boundary conditions \eqref{BC Bondi gauge}, one also imposes the following falloff conditions when $u \to \pm \infty$ that are compatible with the action of superrotations \cite{Compere:2018ylh , Campiglia:2020qvc, Compere:2020lrt, Campiglia:2021bap}:
\begin{equation}
    N_{AB} = N_{AB}^{vac} + o(u^{-2}) ,\qquad C_{AB} = (u + C_\pm) N_{AB}^{vac} - 2 (D_A D_B C_\pm)^{TF} + o(u^{-1}),
    \label{falloff in u}
\end{equation} where $C_{\pm}$ correspond to the values of the supertranslation field at $\mathscr{I}^+_\pm$ that encodes the displacement memory effect \cite{Strominger:2014pwa}. We have the transformation 
\begin{equation}
    \delta_{(f,Y)} C_{\pm} = (\Omega_S \bar{\Omega}_S)^{-\frac{1}{2}} \mathcal{T} + Y^A \partial_A C_{\pm} - \frac{1}{2} D_A Y^A C_\pm .
\end{equation} As discussed in \cite{Campiglia:2020qvc, Campiglia:2021bap}, the falloffs \eqref{falloff in u} are stronger than those considered in e.g. \cite{Compere:2018ylh}, but we found that they are necessary for the finiteness of the flux related to superrotations that we will introduce in Section \ref{Generators and momenta}. 
The falloffs \eqref{falloff in u} imply that, at the corners $\mathscr{I}^+_\pm$, the spacetime is non-radiative ($\hat{N}_{AB}|_{\mathscr{I}^+_\pm} = 0$) and the physical asymptotic shear defined by $\hat{C}_{AB} = C_{AB} - u N^{vac}_{AB}$ is purely electric, i.e. 
\begin{equation}
    \left[\left(D_B D_C - \frac{1}{2} N_{BC}^{vac} \right) \hat{C}_A^C - \left( D_A D_C - \frac{1}{2} N_{AC}^{vac} \right) \hat{C}_B^C \right]\Big|_{\mathscr{I}^+_\pm} = 0. 
    \label{electricity condition}
\end{equation} One can check that this condition is preserved under BMS transformations. It generalizes the standard electricity condition considered e.g. in \cite{Strominger:2013jfa} in presence of superrotations \cite{Compere:2018ylh , Campiglia:2020qvc, Compere:2020lrt}.

\section{Conformal fields on the celestial Riemann surface}
\label{Conformal fields on the celestial Riemann surface}

We now set up the stage for the boundary side of the celestial holography framework. From the previous section, we infer that the celestial Riemannian surface $\mathcal{S} \simeq \mathscr{I}^+ / \mathbb{R}$ can be taken as the $2$-punctured Riemann sphere endowed with the fixed Euclidian metric \eqref{2 sphere metric} \cite{Barnich:2021dta}. It is convenient to complexify $\mathcal{S}$ and treat the coordinates $z$ and $\bar{z}$ independently. From the boundary point of view, one can consider the ``extended conformal transformations'' that preserve the conformal class of the metric \eqref{2 sphere metric}. They are defined as the combined action of conformal coordinate transformations $z'=z'(z)$ and $\bar{z}'=\bar{z}'(\bar{z})$ and Weyl rescalings that induce the following transformations on the conformal factor:
\begin{equation}
    (\Omega \bar{\Omega})'(z', \bar{z}') = (\Omega \bar{\Omega}) (z, \bar{z}) \Big( \frac{\partial z'}{\partial z} \Big) \Big( \frac{\partial \bar{z}'}{\partial \bar{z}} \Big) e^{-2E_R (z', \bar{z}')},
\label{extended conformal transfo}
\end{equation} where $E_R(z, \bar{z})$ is the real Weyl rescaling parameter.\footnote{The parametrization of the Weyl rescaling in \eqref{extended conformal transfo} is precisely the one considered in \cite{Barnich:2016lyg , Freidel:2021fxf , Barnich:2021dta}. It is related to the parametrization used in \cite{Barnich:2010eb, Barnich:2019vzx} through a redefinition of the parameters.} These transformations preserve the particular representative \eqref{2 sphere metric} of the conformal class provided
\begin{equation}
     e^{E_R (z', \bar{z}')}  = \frac{(1+ z \bar{z})}{(1+ z' \bar{z}')} \sqrt{ \Big( \frac{\partial z'}{\partial z} \Big) \Big( \frac{\partial \bar{z}'}{\partial \bar{z}} \Big)} .
     \label{metric preserving}
\end{equation} Infinitesimally, the extended conformal transformations \eqref{extended conformal transfo} satisfying \eqref{metric preserving} are generated by the conformal Killing vectors $\mathcal{Y}(z) \partial+ \bar{\mathcal{Y}}(\bar{z}) \bar{\partial}$ induced on $\mathcal{S}$ from the bulk superrotations defined in \eqref{AKV Bondi}. 

A conformal field of weights $(h, \bar{h})$ is defined as a field $\phi_{h, \bar{h}}(x)$ on $\mathcal{S}$ which transforms as 
\begin{equation}
        \phi'_{h, \bar h} (x') = \Big( \frac{\partial z}{\partial z'} \Big)^h \Big( \frac{\partial \bar z}{\partial \bar z'} \Big)^{\bar h} \phi_{h, \bar h} (x)
        \label{def conformal field}
    \end{equation} under transformations \eqref{extended conformal transfo} with the constraint \eqref{metric preserving}. One can then define a spin weight $J$ and a boost weight/conformal dimension $\Delta$ as usual:
    \begin{equation}
        J = h - \bar{h}, \qquad  \Delta  =  (h+ \bar{h}).
    \end{equation} As explained in \cite{Barnich:2016lyg,Barnich:2021dta}, there is a one-to-one map between conformal fields and weighted scalars with spin weight $s = J$ and boost weight $w=-\Delta$. The weighted scalar point of view is the one that naturally arises when starting from the solution space of gravity \cite{Newman:1961qr,Newman:1966ub,Held:1970kr}. However, it can be easily related to the conformal field point of view by using the conformal factor of the metric. In the present paper, we choose to work in the latter framework which is more adapted to celestial holography.

It will turn out to be useful to introduce the following derivative operators:
\begin{equation}
    \begin{split}
       &\mathscr{D}\phi_{h, \bar{h}} = [D_z - h \partial \Phi ] \phi_{h, \bar{h}},  \qquad 
       \bar{\mathscr{D}}\phi_{h, \bar{h}} = [D_{\bar{z}} - \bar{h} \bar{\partial} \Phi ] \phi_{h, \bar{h}},
    \end{split}  \label{derivative operators conformal}
\end{equation} which act on $(h, \bar{h})$ conformal fields to give conformal fields of weights $(h+1, \bar{h})$ and $(h, \bar{h}+1)$, respectively. They satisfy $\mathscr{D}(\Omega_S\bar{\Omega}_S) = 0 = \bar{\mathscr{D}}(\Omega_S\bar{\Omega}_S)$. These operators coincide with the Weyl covariant derivative operators introduced in \cite{Barnich:2021dta} for the framework that we are considering here; i.e. with fixed representative \eqref{2 sphere metric}. In particular, the Liouville field introduced in \eqref{Liouville stress tensor} naturally arises as part of the Weyl connection. Furthermore, the operators \eqref{derivative operators conformal} also correspond to the $\text{Witt}\oplus\overline{\text{Witt}}$ version of the Diff$(S^2)$-covariant derivative introduced in \cite{Campiglia:2020qvc} (see also \cite{Campiglia:2021bap}). One can check that \begin{equation}
    [\mathscr{D}, \bar{\mathscr{D}}] \phi_{h, \bar{h}} = 0.
\end{equation}
    
We assume that the conformal fields can be expanded in formal series as
\begin{equation}
    \phi_{h, \bar h} (z, \bar{z})= \sum_{k,l} \,  a_{k,l} \, {}_{h,\bar h}Z_{k,l} , \qquad {}_{h,\bar h}Z_{k,l} = z^{-h-k} \bar{z}^{-\bar h - l},
    \label{series expansion}
\end{equation} where the coefficients $a_{k,l} \in \mathbb C$ satisfy appropriate conditions \cite{book:4760 , book:75861}. If $h,\bar h$ are integers (resp. half integers), then $k,l$ are taken to be integers (resp. half integers). The residues of $\phi_{h,\bar h}(z, \bar{z})$ with respect to $z$ and $\bar{z}$ are defined as
\begin{equation}
    Res_{z}[\phi_{h, \bar{h}}(z, \bar{z})] = \sum_{l} a_{1-h, l} \bar{z}^{-\bar{h}-l}, \qquad Res_{\bar{z}}[\phi_{h, \bar{h}}(z, \bar{z})] = \sum_{k} a_{k, 1-\bar{h}} {z}^{-{h}-k}.
\end{equation} From the residue theorem, we have the fundamental relations
\begin{equation}
    \oint_{\mathcal{C}} \frac{dz}{2i\pi} \, \phi_{h, \bar{h}}(z, \bar{z}) = Res_{z}[\phi_{h, \bar{h}}(z, \bar{z})], \qquad \oint_{\mathcal{C}} \frac{d\bar{z}}{2i\pi} \, \phi_{h, \bar{h}}(z, \bar{z}) = Res_{\bar{z}}[\phi_{h, \bar{h}}(z, \bar{z})],
\end{equation} where $\mathcal{C}$ is a contour around the puncture. Notice that total derivative terms can be discarded in the contour integrals since there is no $\log z$ terms in the expansion \eqref{series expansion}. We use the notation $\int_{\mathcal{S}} \frac{dzd\bar{z}}{(2i \pi)^2} = \oint_{\mathcal{C}} \frac{dz}{2i\pi}\int_{\mathcal{C}}\frac{d\bar{z}}{2i\pi}$ to designate the integral over the celestial Riemann surface $\mathcal{S}$.\footnote{\label{footnore ref measure}Notice that the standard normalization for the measure on the celestial sphere is such that \[ \int_{\mathcal{S}} idz d\bar{z} \, (\Omega_S \bar{\Omega}_S)^{-1} = \int^\pi_0 d\theta \, \sin^2 \theta  \int^{2\pi}_0 d\phi \,  = 4 \pi \] when using stereographic coordinates $z = \cot \left(\frac{\theta}{2} \right) e^{-i\phi}$, $\bz = \cot \left(\frac{\theta}{2} \right) e^{i\phi}$.} The expansion \eqref{series expansion} is inverted by the relation 
\begin{equation}
    a_{k,l} = \int_{\mathcal{S}} \frac{dzd\bar{z}}{(2i\pi)^2} \, z^{k+h-1} \bar{z}^{l+\bar{h}-1} \phi_{h,\bar{h}}(z, \bar{z}).
\end{equation} The formal Dirac delta-functions are defined as the following formal distributions \cite{book:4760 , book:75861}: 
\begin{equation}
  \delta (z -w) = z^{-1}\sum_{k\in \mathbb{Z}} \left( \frac{z}{w} \right)^k, \qquad \delta (\bar{z}-\bar{w}) =  \bar{z}^{-1}\sum_{k\in \mathbb{Z}} \left( \frac{\bar{z}}{\bar{w}} \right)^k.
\end{equation} They satisfy the useful properties $\oint_{\mathcal{C}} \frac{dz}{2i\pi} \delta (z-w) \phi_{h, \bar{h}}(z, \bar{z}) = \phi_{h, \bar{h}}(w, \bar{z})$, $\delta (z-w) = \delta (w-z)$, $(z-w)\delta (z-w)=0$, $\partial_z \delta (z-w) = - \partial_w \delta (z-w)$, together with the corresponding relations for $\delta (\bar{z} - \bar{w})$. We write the delta-function on the celestial Riemann surface as 
\begin{equation}
    \delta^2 (z-w) = \delta (z-w) \delta (\bar{z}- \bar{w}),
\end{equation} so that $
    \int_{\mathcal{S}} \frac{dz d\bar{z}}{(2i\pi)^2} \, \phi_{h, \bar{h}}(z, \bar{z}) \delta^2 (z-w) = \phi_{h, \bar{h}}(w, \bar{w})$.

\section{Generators and momenta}
\label{Generators and momenta}

We now identify the parameters of the extended BMS algebra and some non-local combinations of the solution space introduced in Section \ref{Asymptotically flat spacetimes} from bulk considerations, as conformal fields on the celestial Riemann surface in the sense of Section \ref{Conformal fields on the celestial Riemann surface}. The conformal weights of the various fields discussed in this paper are summarized in Table \ref{weights S}. 

Let us start with the $\mathfrak{bms}_4$ symmetry parameter given in Equation \eqref{AKV Bondi}.\footnote{In this paper, since we are focusing on the conformal field point of view, we do not write the `` $\tilde{}$ '' notation above the conformal fields, which contrasts with the convention used in \cite{Barnich:2021dta}.} The supertranslation parameters ${\mathcal T}(z,\bar{z})$ can be seen as real conformal fields  of weights $(-\frac{1}{2},-\frac{1}{2})$ \cite{Barnich:2021dta}.\footnote{The choice of conformal weights for the symmetry parameters will be justified later through the pairing between generators and momenta in \eqref{BMS flux}.} They can be expanded in formal series as in  \eqref{series expansion}:
\begin{equation}
    \mathcal{T}(z, \bar{z}) = \sum_{k,l} t_{k,l} \mathscr{T}_{k,l}, \qquad \mathscr{T}_{k,l} = {}_{-\frac{1}{2},-\frac{1}{2}}
   Z_{k,l}=z^{\frac{1}{2}-k}
  \bar{z}^{\frac{1}{2}-l},
  \label{commu 1 prime}
\end{equation} with $k,l$ half-integers and $\bar{t}_{k,l} = t_{l,k}$ so that $\mathcal{T}$ is real. Notice that the four Poincaré translations are spanned by $\mathscr{T}_{\12,\12}$, $\mathscr{T}_{\12,-\12}$, $\mathscr{T}_{-\12,\12}$ and $\mathscr{T}_{-\12,-\12}$.

Superrotations are parametrized by the complex conformal fields $\mathcal{Y}(z)$, $\bar{ \mathcal{Y}}(\bar z)$ of weights $(-1,0)$ and $(0,-1)$, respectively \cite{Barnich:2021dta}. The infinitesimal actions of superrotations on a conformal field $\phi_{h,\bar{h}}(z, \bar{z})$ are given by 
\begin{equation}
    \delta_\mathcal{Y}  \phi_{h, \bar h} =  \mathcal{Y} \partial \phi_{h, \bar h} + h \partial \mathcal{Y} \phi_{h, \bar h} , \qquad \delta_{\bar{\mathcal{Y}}} \phi_{h, \bar h} = \bar{\mathcal{Y}} \bar{\partial} \phi_{h, \bar h} + \bar{h} \bar{\partial} \bar{\mathcal{Y}} \phi_{h, \bar h}.
    \label{commu 2}
\end{equation} These are the infinitesimal analogues of \eqref{def conformal field}. Superrotation vector fields can be expanded as in \eqref{series expansion}:
\begin{equation}
   \mathcal{Y}(z) = \sum_{m} y_m \mathcal{Y}_m, \quad \mathcal{Y}_m = z^{1-m}, \qquad \bar{\mathcal{Y}}(\bar{z}) = \sum_{m} \bar{y}_m \bar{\mathcal{Y}}_m, \quad \bar{\mathcal{Y}}_m = \bar{z}^{1-m},
    \label{commu 2 prime}
\end{equation} where $m \in \mathbb{Z}$ and $y_m, \bar{y}_m$ are complex numbers. The six global Lorentz parameters are spanned by $\mathcal{Y}_{-1}$, $\mathcal{Y}_{0}$, $\mathcal{Y}_{1}$, and their complex conjugates. In terms of \eqref{commu 2 prime}, equation \eqref{commu 2} can be rewritten as
\begin{equation}
   \delta_{\mathcal{Y}_{m}}  \phi_{h, \bar h} =  z^{1-m} \partial \phi_{h, \bar h} + h (1-m) z^{-m} \phi_{h, \bar h}, \qquad  \delta_{\bar{\mathcal{Y}}_{m}}  \phi_{h, \bar h} =  \bar{z}^{1-m} \bar{\partial} \phi_{h, \bar h} + \bar{h} (1-m) \bar{z}^{-m} \phi_{h, \bar h}
\end{equation} or, equivalently,
\begin{equation}
    \delta_{\mathcal{Y}_m}  ({}_{h,\bar{h}}Z_{k,l}) = -( k + h m) \,{}_{h, \bar{h}} Z_{m+k, l}, \qquad \delta_{\bar{\mathcal{Y}}_m} ({}_{h,\bar{h}}Z_{k,l}) = - (l+\bar{h}m)  \,{}_{h,  \bar{h}}Z_{k, l+m}.
     \label{commu 2 prime 2}
\end{equation} For convenience, the $\mathfrak{bms}_4$ commutation relations \eqref{commutation relations 1} and \eqref{commutation relations 2} can be rewritten using the notations $\xi(\mathcal{T},0,0) \to \mathcal{T}$, $\xi(0, \mathcal{Y}, 0) \to \mathcal{Y}$, $\xi (0,0, \bar{\mathcal{Y}}) \to \bar{\mathcal{Y}}$, $[.,.]_\star \to [.,.]$. We have
\begin{equation}
\begin{split}
&[\mathcal{Y}_1, \mathcal{Y}_2] = \mathcal{Y}_1 \partial \mathcal{Y}_2 - \mathcal{Y}_2 \partial \mathcal{Y}_1, \qquad [\bar{\mathcal{Y}}_1, \bar{\mathcal{Y}}_2] = \bar{\mathcal{Y}}_1 \bar{\partial} \bar{\mathcal{Y}}_2 - \bar{\mathcal{Y}}_2 \bar{\partial} \bar{\mathcal{Y}}_1,\qquad [\mathcal{T}_1, \mathcal{T}_2] = 0,\\ 
&[\mathcal{Y}_1 , \mathcal{T}_2] =  \mathcal{Y}_1 \partial \mathcal{T}_2 - \frac{1}{2} \partial \mathcal{Y}_1 \mathcal{T}_2 , \qquad [\bar{\mathcal{Y}}_1, \mathcal{T}_2] = \bar{\mathcal{Y}}_1 \bar{\partial} \mathcal{T}_2 - \frac{1}{2} \bar{\partial}\bar{\mathcal{Y}}_1 \mathcal{T}_2.
\end{split}
\end{equation} In terms of expansions \eqref{commu 1 prime} and \eqref{commu 2 prime}, these commutation relations become \cite{Barnich:2010eb,Barnich:2017ubf}
\begin{equation}
  \begin{split}
   &[ {\mathcal{Y}}_m, {\mathcal{Y}}_n ] = (m-n) {\mathcal{Y}}_{m+n} , \quad [
    {\bar{\mathcal{Y}}}_m, {\bar{\mathcal{Y}}}_n ] = (m-n){\bar{\mathcal{Y}}}_{m+n}, \\ 
    &[{\mathcal{Y}}_m ,
\mathscr{T}_{k,l} ] = \Big( \frac{1}{2} m - k \Big)
\mathscr{T}_{m+k,l}, \quad [{\bar{\mathcal{Y}}}_m , \mathscr{T}_{k,l} ] =
\Big(\frac{1}{2} m - l\Big) \mathscr{T}_{k,m+l}, \\ &[ {\mathcal{Y}}_m ,
{\bar{\mathcal{Y}}}_n ] = 0 = [ \mathscr{T}_{k,l} ,
\mathscr{T}_{r,s}] .
\end{split} 
\end{equation}

Let us now define the BMS momentum fluxes as particular non-local combinations of the solution space data of asymptotically flat spacetimes appearing in \eqref{fall-off} and interpret them as conformal fields on the celestial Riemann surface $\mathcal{S}$. 

First, the \emph{supermomentum flux} $\mathcal{P}(z, \bar z)$
is defined by 
\begin{equation}
        \mathcal{P} = \frac{1}{4\pi G} \int^{+\infty}_{-\infty} du\, \partial_u \mathcal{M} = \frac{\mathcal{M}}{4\pi G}\Big|^{\mathscr{I}^+_+}_{\mathscr{I}^+_-}, \qquad
        \mathcal{M} = (\Omega_S \bar{\Omega}_S)^{-\frac{3}{2}} \Big[ M + \frac{1}{8} ( C_{zz} N^{zz}_{vac}+ C_{\bar{z}\bar{z}} N^{\bar{z}\bar{z}}_{vac})\Big],
        \label{supermomentum charge}
    \end{equation} which corresponds to the difference of values of the supermomentum $\mathcal{M}(u,z,\bar{z})$ between the two non-radiative asymptotic regions $\mathscr{I}^+_\pm$. Here we used the prescription of \cite{Compere:2018ylh, Compere:2020lrt} to define the supermomentum in terms of the solution space. The supermomentum flux $\mathcal{P}(z, \bar{z})$ can be seen as a ($J=0$) conformal field of weights $(\frac{3}{2},\frac{3}{2})$. Indeed, under infinitesimal BMS transformations, one can deduce from \eqref{transformation on the solution space} and \eqref{falloff in u} that
    \begin{equation}
      \delta_{(\mathcal{T}, \mathcal{Y}, \bar{\mathcal{Y}})} \mathcal{P} = \Big[\mathcal{Y} \partial + \bar{\mathcal{Y}} \bar{\partial} + \frac{3}{2} \partial \mathcal{Y} + \frac{3}{2} \bar{\partial} \bar{\mathcal{Y}} \Big] \mathcal{P},
        \label{coadjoint P}
    \end{equation}  which is the expected infinitesimal transformation law for a $(\frac{3}{2},\frac{3}{2})$ conformal field (see \eqref{commu 2}). For later purposes, it is useful to split the supermomentum flux into soft and hard parts involving linear, respectively quadratic, terms in $\hat{N}_{AB}$ inside the integral. We prescribe
    \begin{equation}
        \mathcal{P} =  \mathcal{P}_{soft} + \mathcal{P}_{hard}  
    \end{equation} with
    \begin{equation}
        \begin{split}
 \mathcal{P}_{soft} &= \frac{1}{16 \pi G}  \int^{+\infty}_{-\infty} du \, (\Omega_S \bar{\Omega}_S)^{-\frac{3}{2}}\Big[ \Big(D_z^2 - \frac{1}{2} N_{zz}^{vac} \Big)  \hat{N}^{zz} + \Big( D_{\bar{z}}^2 - \frac{1}{2} N^{vac}_{\bar{z}\bar{z}} \Big) \hat{N}^{{\bar{z}}{\bar{z}}}  \Big], \\
 \mathcal{P}_{hard} &= - \frac{1}{16\pi G}  \int^{+\infty}_{-\infty} du \, (\Omega_S \bar{\Omega}_S)^{-\frac{3}{2}}\Big[ \hat{N}_{zz} \hat{N}^{zz} \Big].
 \label{soft hard P}
        \end{split}
    \end{equation} Notice that the fluxes of supermomenta defined as in \eqref{soft hard P} are finite in $u$ and vanish in stationary configurations where $\hat{N}_{AB} = 0$, which is a desirable physical requirement \cite{Wald:1999wa, Compere:2018ylh}. One can show that the soft and the hard parts transform separately as in \eqref{coadjoint P}, namely
    \begin{equation}
      \delta_{(\mathcal{T}, \mathcal{Y}, \bar{\mathcal{Y}})} \mathcal{P}_{soft/hard} = \Big[\mathcal{Y} \partial + \bar{\mathcal{Y}} \bar{\partial} + \frac{3}{2} \partial \mathcal{Y} + \frac{3}{2} \bar{\partial} \bar{\mathcal{Y}} \Big]  \mathcal{P}_{soft/hard}.
        \label{coadjoint P soft part}
    \end{equation} They can therefore be both interpreted as conformal fields of weights $(\frac{3}{2}, \frac{3}{2})$, which justifies the specific split between hard and soft parts in \eqref{soft hard P}. 
     In terms of the superrotation-covariant derivative operators introduced in \eqref{derivative operators conformal}, the soft part can be elegantly rewritten as 
    \begin{equation}
     \mathcal{P}_{soft} = \mathscr{D}^2 \bar{\mathscr{N}}^{(0)} + \bar{\mathscr{D}}^2 \mathscr{N}^{(0)}, \qquad \mathscr{N}^{(0)} = \frac{1}{16 \pi G} \int^{+\infty}_{-\infty}  du \, (\Omega_S\bar{\Omega}_S)^{\frac{1}{2}}  \hat{N}_{{{z}}{{z}}} ,
    \end{equation} where the leading soft mode of the news tensor $\mathscr{N}^{(0)}(z, \bar{z})$ is a $(\frac{3}{2} ,-\frac{1}{2})$ conformal field. Moreover, using the electricity condition \eqref{electricity condition} encoded in the falloffs \eqref{falloff in u}, we have $\mathscr{D}^2 \bar{\mathscr{N}}^{(0)} = \bar{\mathscr{D}}^2 \mathscr{N}^{(0)}$ and 
    \begin{equation}
    \begin{split}
       \mathcal{P}_{soft} &= 2 \mathscr{D}^2 \bar{\mathscr{N}}^{(0)} \\
       &= \frac{1}{8 \pi G}(\Omega_S\bar{\Omega}_S)^{\frac{1}{2}} \Big(D_z^2 - \frac{1}{2} N_{zz}^{vac} \Big)  [ \Delta C N_{\bar{z}\bar{z}}^{vac} - 2 D^2_{\bar{z}} \Delta C ],
       \label{integrated expression P}
       \end{split}
    \end{equation} with $\Delta C = C_{+} - C_{-}$ the difference of the supertranslation field between the future and past corners of $\mathscr{I}^+$.

   Second, the \emph{super angular momentum flux} is parametrized by the equivalence classes $[\mathcal{J}]$ and $[\bar{\mathcal{J}}]$ for the equivalence relation \cite{Barnich:2021dta}
    \begin{equation}
        \mathcal{J} \equiv \mathcal{J} + \mathscr{D} \mathcal{L} , \qquad  \bar{\mathcal{J}} \equiv \bar{\mathcal{J}} + \bar{\mathscr{D}} \bar{\mathcal{L}}.
        \label{equivalence classes}
    \end{equation} These equivalence classes can be seen as complex conformal fields of weights $(1,2)$ and $(2,1)$, respectively. In terms of the gravitational data, we define
    \begin{equation}
    \begin{split}
        \mathcal{J} = \frac{1}{8\pi G} \int^{+\infty}_{-\infty} du \, \partial_u \mathcal{N} = \frac{\mathcal{N}}{8\pi G}\Big|^{\mathscr{I}^+_+}_{\mathscr{I}^+_-},
        \end{split}
        \end{equation} which corresponds to the difference of values between $\mathscr{I}^+_{\pm}$ of the super angular momentum $\mathcal{N}(u,z, \bar{z})$ defined by
        \begin{equation}
        \begin{split}
        \mathcal{N} &= (\Omega_S \bar{\Omega}_S)^{-1} \Big[ N_{\bar{z}} - u \Omega_S^3 D_{\bar{z}}  \mathcal{M} + \frac{1}{4} C_{\bar{z}\bar{z}} D_{\bar{z}} C^{\bar{z}\bar{z}} + \frac{3}{16} D_{\bar{z}} (C_{zz}C^{zz}) \Big]\\
        &\quad+ \frac{u}{4}(\Omega_S \bar{\Omega}_S)^{-1} D^z \Big[ \Big( D_z^2 - \frac{1}{2} N^{vac}_{zz}   \Big)C^z_{\bar{z}} - \Big( D_{\bar{z}}^2 - \frac{1}{2} N^{vac}_{\bar{z}\bar{z}}\Big)C^{\bar{z}}_z \Big].
    \end{split}
    \label{definition super angular momentum}
    \end{equation} We have the analogous complex conjugate relations for $\bar{\mathcal{J}}(z, \bar{z})$ and $\bar{\mathcal{N}}(u,z,\bar{z})$. Here we used a prescription based on \cite{Compere:2018ylh ,Compere:2020lrt,Compere:2021inq} to define the super angular momentum.\footnote{\label{reffoofnote}The prescription \eqref{definition super angular momentum} for the super angular momentum differs from the one proposed in \cite{Compere:2020lrt} by magnetic contributions of the shear that do not play any role at $\mathscr{I}^+_{\pm}$, but that allow to have vanishing fluxes for stationary solutions $\hat{N}_{AB} = 0$ (see \cite{AdrienThesis} for a detailed discussion).} Under infinitesimal BMS transformations acting through \eqref{transformation on the solution space}, the super angular momentum flux transforms as
    \begin{equation}
       \delta_{(\mathcal{T}, \mathcal{Y}, \bar{\mathcal{Y}})} \mathcal{J} = \mathcal{Y} \partial \mathcal{J} + \bar{\mathcal{Y}} \bar{\partial} \mathcal{J}+ \partial  \mathcal{Y}\mathcal{J} + 2 \bar{\partial} \bar{\mathcal{Y}} \mathcal{J}   + \frac{1}{2} \mathcal{T} \bar{\partial} \mathcal{P} + \frac{3}{2} \bar{\partial} \mathcal{T} \mathcal{P},
       \label{coadjoint J}
    \end{equation} together with the complex conjugate relation for $\bar{\mathcal{J}}$. For future purposes, it is useful to split the super angular momentum flux into soft and hard parts; we propose
    \begin{equation}
        \mathcal{J} =  \mathcal{J}_{soft} + \mathcal{J}_{hard}  
    \end{equation} with
    \begin{equation}
        \begin{split}
          \mathcal{J}_{soft} &= - \frac{1}{16\pi G} \int^{+\infty}_{-\infty} du \, (\Omega_S \bar{\Omega}_S)^{-1} \Big[ u \Big( D_{\bar{z}}^3 - 2 N_{\bar{z}\bar{z}}^{vac} {D}_{\bar{z}} -  D_{\bar{z}} N^{vac}_{\bar{z}\bar{z}}  \Big) \hat{N}^{\bar{z}\bar{z}}  \\ &\qquad\qquad\qquad\qquad+\hat{N}^{\bar{z}\bar{z}} D_{\bar{z}}\Big[\Big(D^2_{\bar{z}} - \frac{1}{2}  N_{\bar{z}\bar{z}}^{vac}  \Big) C_-\Big]+ 3 D_{\bar{z}} \hat{N}^{\bar{z}\bar{z}} \Big( D_{\bar{z}}^2 - \frac{1}{2} N_{\bar{z}\bar{z}}^{vac}  \Big) C_- \Big], \\
          \mathcal{J}_{hard} &= \frac{1}{8\pi G} \int^{+\infty}_{-\infty} du \, (\Omega_S \bar{\Omega}_S)^{-1} \Big[\frac{3}{4}\hat{C}_{\bar{z}\bar{z}} D_{\bar{z}} \hat{N}^{\bar{z}\bar{z}} + \frac{1}{4} \hat{N}^{\bar{z}\bar{z}} D_{\bar{z}} \hat{C}_{\bar{z}\bar{z}} +\frac{u}{4} D_{\bar{z}} ( \hat{N}_{\bar{z}\bar{z}} \hat{N}^{\bar{z}\bar{z}}  )  \\ &\qquad\qquad\qquad\qquad+\frac{1}{2}\hat{N}^{\bar{z}\bar{z}} D_{\bar{z}}\Big[\Big(D^2_{\bar{z}} - \frac{1}{2}  N_{\bar{z}\bar{z}}^{vac}  \Big) C_-\Big]+ \frac{3}{2} D_{\bar{z}} \hat{N}^{\bar{z}\bar{z}} \Big( D_{\bar{z}}^2 - \frac{1}{2} N_{\bar{z}\bar{z}}^{vac}  \Big) C_- \Big]
           \label{hard soft J}
        \end{split}
    \end{equation} and the complex conjugate relations for $\bar{\mathcal{J}}_{soft}$ and $\bar{\mathcal{J}}_{hard}$. Notice that the fluxes of super angular momenta \eqref{hard soft J} are finite thanks to the stronger $u$-falloffs that were taken \eqref{falloff in u} and vanish in stationary configurations where $\hat{N}_{AB} = 0$ \cite{Wald:1999wa, Compere:2018ylh}. One can show that the soft and the hard parts transform separately as in \eqref{coadjoint J}, namely
    \begin{equation}
        \begin{split}
    \delta_{(\mathcal{T}, \mathcal{Y}, \bar{\mathcal{Y}})} \mathcal{J}_{soft} &= \mathcal{Y} \partial \mathcal{J}_{soft} + \bar{\mathcal{Y}} \bar{\partial} \mathcal{J}_{soft} + 2 \bar{\partial} \bar{\mathcal{Y}} \mathcal{J}_{soft} + \partial  \mathcal{Y}\mathcal{J}_{soft}  + \frac{1}{2} \mathcal{T} \bar{\partial} \mathcal{P}_{soft} + \frac{3}{2} \bar{\partial} \mathcal{T} \mathcal{P}_{soft},\\
    \delta_{(\mathcal{T}, \mathcal{Y}, \bar{\mathcal{Y}})} \mathcal{J}_{hard} &= \mathcal{Y} \partial \mathcal{J}_{hard} + \bar{\mathcal{Y}} \bar{\partial} \mathcal{J}_{hard} + 2 \bar{\partial} \bar{\mathcal{Y}} \mathcal{J}_{hard} + \partial  \mathcal{Y}\mathcal{J}_{hard}  + \frac{1}{2} \mathcal{T} \bar{\partial} \mathcal{P}_{hard} + \frac{3}{2} \bar{\partial} \mathcal{T} \mathcal{P}_{hard},
        \end{split}
        \label{transfo soft part J}
    \end{equation} which implies that $\mathcal{J}_{soft}$ and $\mathcal{J}_{hard}$ ($\bar{\mathcal{J}}_{soft}$ and $\bar{\mathcal{J}}_{hard}$) can be seen separately as conformal fields of weights $(1,2)$ (respectively $(2,1)$). The transformations \eqref{transfo soft part J} justify the specific choice of split prescribed between soft and hard parts in \eqref{hard soft J}. In particular, the terms involving the supertranslation field $C_-$ at $\mathscr{I}^+_-$ ensure that the expressions transform as they should under supertranslations.\footnote{One could have replaced $C_-$ by $C_+$ in the expressions \eqref{hard soft J} without affecting the result \eqref{transfo soft part J}.} In terms of the derivative operators \eqref{derivative operators conformal}, the soft part can be rewritten as 
    \begin{equation}
    \begin{split}
    &\mathcal{J}_{soft} =- \bar{\mathscr{D}}^3 \mathscr{N}^{(1)} -\bar{\mathscr{D}}^3  \mathscr{C} \mathscr{N}^{(0)} - 3 \bar{\mathscr{D}}^2 \mathscr{C} \bar{\mathscr{D}} \mathscr{N}^{(0)} , \\
    &\mathscr{N}^{(1)} = \frac{1}{16 \pi G}  \int^{+\infty}_{-\infty}  du \, (\Omega_S\bar{\Omega}_S) u  \hat{N}_{{{z}}{{z}}} , \qquad \mathscr{C} = (\Omega_S \bar{\Omega}_S)^{\frac{1}{2}} C_-,
    \end{split}
    \end{equation} where the subleading soft mode of the news tensor $\mathscr{N}^{(1)}(z, \bar{z})$ is a $(1, -1)$ conformal field and $\mathscr{C}(z, \bar{z})$ is a $(-\frac{1}{2}, -\frac{1}{2})$ conformal field.

    As already mentioned, the prescription for the BMS momenta \eqref{supermomentum charge} and \eqref{definition super angular momentum} that we are using here have all the desired properties, including finiteness in $u$, vanishing for non-radiative solutions, vanishing for vacuum configurations and closure under the standard bracket when considering the integrated fluxes over $\mathscr{I}^+$ (see Section \ref{sec:BMS flux algebra}). Concerning the splits between soft and hard parts in \eqref{soft hard P} and \eqref{hard soft J}, they differ from those originally proposed in \cite{Strominger:2013jfa, He:2014laa, Kapec:2014opa} (see also \cite{Strominger:2017zoo}) by terms involving the memory fields $\varphi(z)$, $\bar{\varphi}(\bar{z})$ and $C_-$ that label the vacuum degeneracy \cite{Compere:2016jwb , Compere:2016hzt , Compere:2016gwf}. When setting $\varphi = 0 = \bar{\varphi}$ ($N_{AB}^{vac}=0$) and $C_-= 0$, we consistently recover the standard expressions. The additional terms that we have allow us to obtain better transformation laws \eqref{coadjoint P soft part} and \eqref{transfo soft part J} under the action of BMS symmetries. As we will see in Section \ref{Relation with supertranslation operator and stress-tensor}, this will be of major importance to identify the CCFT operators in the solution space of gravity that obey the desired constraints and transformation properties.

\section{BMS flux algebra}
\label{sec:BMS flux algebra}

When using covariant phase space methods \cite{Lee:1990nz,Wald:1993nt,Iyer:1994ys,Barnich:2001jy}, BMS surface charges are non-integrable due to the presence of radiation \cite{Wald:1999wa,Barnich:2011mi}. Defining meaningful finite charges requires to impose additional criteria to isolate a specific integrable part (see e.g. \cite{Flanagan:2015pxa,Compere:2018ylh,Harlow:2019yfa,Freidel:2020xyx,Compere:2020lrt,Freidel:2021fxf,Freidel:2021cbc} for recent proposals of such criteria and \cite{Compere:2019gft, Elhashash:2021iev} for the implication of the various expressions on observational data). Here, we follow the prescription of \cite{Compere:2020lrt,Compere:2021inq,AdrienThesis} (see also footnote \ref{reffoofnote}) to select the integrable part and define the ``finite'' charges. The fluxes are then obtained by expressing the finite surface charge integrals as volume integrals over $\mathscr{I}^+$. In terms of the generators and flux of momenta that were introduced in Section \ref{Generators and momenta}, the BMS fluxes read as\footnote{Notice that to obtain the standard normalization for the BMS fluxes, one should multiply the expression \eqref{BMS flux} by a global factor $-4i \pi^2$ (see also footnote \ref{footnore ref measure}).}  
\begin{equation}
    F_{(\mathcal{T}, \mathcal{Y}, \bar{\mathcal{Y}})} =  \int_{\mathcal{S}} \frac{dz d\bar{z}}{(2i\pi)^2} \, [\mathcal{T} \mathcal{P} + \mathcal{Y} \bar{\mathcal{J}} + \bar{\mathcal{Y}} \mathcal{J} ] = \frac{1}{8\pi G} \int_{\mathscr{I}^+} du \frac{dz d\bar{z}}{(2i\pi)^2}\, \partial_u [2\mathcal{T} \mathcal{M} + \mathcal{Y} \bar{\mathcal{N}} + \bar{\mathcal{Y}} \mathcal{N}].
    \label{BMS flux}
\end{equation} As mentioned above, the conformal weights of the BMS parameters $(\mathcal{T}, \mathcal{Y}, \bar{\mathcal{Y}})$ are completely determined by those of the measure $dzd\bar{z}$ and the momentum fluxes $(\mathcal{P},[\mathcal{J}], [\bar{\mathcal{J}}])$ by requiring that the total flux $F_{(\mathcal{T}, \mathcal{Y}, \bar{\mathcal{Y}})}$ has vanishing conformal weights; see Table \ref{weights S}.

The expression \eqref{BMS flux} can be seen as a pairing $\langle \cdot,\cdot\rangle$ between the BMS generators of the algebra and the momentum fluxes:
\begin{equation}
    \mathfrak{bms}_4^* \times \mathfrak{bms}_4 \mapsto \mathbb R : \left((\mathcal{P}, [\mathcal{J}], [\bar{\mathcal{J}}]) , (\mathcal{T}, \mathcal{Y}, \bar{\mathcal{Y}}) \right) \to  F_{(\mathcal{T}, \mathcal{Y}, \bar{\mathcal{Y}})} = \langle (\mathcal{P}, [\mathcal{J}], [\bar{\mathcal{J}}]) , (\mathcal{T}, \mathcal{Y}, \bar{\mathcal{Y}}) \rangle ,
\end{equation} where $\mathfrak{bms}_4^*$ denotes the dual of $\mathfrak{bms}_4$. Indeed, it is linear in both entries and non-degeneracy comes from the fact that we have considered equivalence classes $[\mathcal{J}]$ and $[\bar{\mathcal{J}}]$ of super angular momentum fluxes \eqref{equivalence classes}. Using this pairing, the transformations laws \eqref{coadjoint P} and \eqref{coadjoint J} can be interpreted as the coadjoint representation of $\mathfrak{bms}_4$ \cite{Barnich:2021dta}. In particular, it has been shown in that reference that there exists a (pre-)momentum map between the solution space space of non-radiative asymptotically flat spacetimes and the dual of the global BMS algebra $\mathfrak{so(3,1)} \loplus \mathfrak{s}$. Here, we have extended these results for radiative spacetimes and for extended BMS algebra  $(\text{Witt} \oplus \overline{\text{Witt}}) \loplus \mathfrak{s}^*$ by considering the fluxes on $\mathscr{I}^+$, which are $u$-integrated expressions in terms of the solution space, rather than surface charges. These results rely crucially on the falloff conditions \eqref{falloff in u} at the corners of $\mathscr{I}^+$ and the fact that the BMS fluxes are determined by the values of the surface charges at the corners. 

Now, using the basis dual to the one used for the expansion of the generators in \eqref{commu 1 prime} and \eqref{commu 2 prime} \cite{Barnich:2021dta}, it is instructive to expand the BMS momentum fluxes as in \eqref{series expansion}:
     \begin{equation}
        \begin{split}
       &\mathcal{P} (z, \bar{z})= \sum_{k,l} p_{k,l} \mathscr{T}_*^{k,l}, \qquad \mathscr{T}_*^{k,l} = {}_{\frac{3}{2},\frac{3}{2}}Z_{-l,-k}  = {z}^{-\frac{3}{2}+l}
{\bar{z}}^{-\frac{3}{2}+k}, \\
 &\mathcal{J}(z, \bar{z})= \sum_{m} {j}_m
  \mathcal{Y}^m_* ,\qquad \mathcal{Y}^m_* = {}_{1,2}Z_{0,-m}=
z^{-1} {\bar{z}}^{-2+m}, \\
&\bar{\mathcal{J}}(z, \bar{z})= \sum_{m}
  {\bar{j}}_m \bar{\mathcal{Y}}_*^m , \qquad \bar{\mathcal{Y}}_*^m =  {}_{2,1}Z_{-m,0} = {z}^{-2+m}
\bar{z}^{-1}.
\end{split}
\end{equation} with $\bar{p}_{k,l} = p_{l,k}$ so that $\mathcal{P}$ is real. In terms of the above expansions, the infinitesimal variations \eqref{coadjoint P} and \eqref{coadjoint J} are encoded in the coadjoint representation of $\mathfrak{bms}_4$, written $\text{ad}^*$, as follows \cite{Barnich:2021dta}:
\begin{equation}
  \begin{split}
    &\text{ad}^*_{\mathcal{Y}_m} {\mathcal{Y}}^n_* = (-2m+n)\mathcal{Y}^{n-m}_*,
    \quad \text{ad}^*_{\bar{\mathcal{Y}}_m}  \bar{\mathcal{Y}}^n_*  = (-2m+n)
    \bar{\mathcal{Y}}_*^{n-m}, \\ &\text{ad}^*_{\mathcal{Y}_m} \mathscr{T}^{k,l}_* = \Big(-
    \frac{3}{2} m +k \Big) \mathscr{T}^{k-m,l}_* , \quad \text{ad}^*_{\bar{\mathcal{Y}}_m} 
    \mathscr{T}^{k,l}_* = \Big( -\frac{3}{2} m + l \Big)
    \mathscr{T}^{k,l-m}_* , \\ &\text{ad}^*_{\mathscr{T}_{k,l}} \mathscr{T}^{r,s}_*
    = \Big(\frac{r-3k}{2}\Big) \delta^s_l\mathcal{Y}_*^{r-k} + \Big( \frac{s-
      3 l}{2} \Big) \delta^r_k\bar{\mathcal{Y}}^{s-l}_* , \\ &\text{ad}^*_{\mathcal{Y}_m} 
    \bar{\mathcal{Y}}^n_*  = 0 = \text{ad}^*_{\bar{\mathcal{Y}}_m}  \bar{\mathcal{Y}}^n_*  , \quad
    \text{ad}^*_{\mathscr{T}_{k,l}}  \mathcal{Y}^m_* = 0 = \text{ad}^*_{\mathscr{T}_{k,l}}
    \bar{\mathcal{Y}}^m_* .\label{coadjoint in conformal basis}
\end{split}
\end{equation}

Similarly, the soft/hard BMS fluxes
\begin{equation}
    F_{(\mathcal{T}, \mathcal{Y}, \bar{\mathcal{Y}})}^{soft/hard} =  \int_{\mathcal{S}} \frac{dz d\bar{z}}{(2i\pi)^2}  \, [\mathcal{T} \mathcal{P}_{soft/hard} + \mathcal{Y} \bar{\mathcal{J}}_{soft/hard} + \bar{\mathcal{Y}} \mathcal{J}_{soft/hard} ] 
    \label{pairing hard soft}
\end{equation} play the role of pairing for the soft/hard sectors. As discussed in Section \ref{Generators and momenta}, since the soft and hard parts of the momentum fluxes transform separately as \eqref{coadjoint P} and \eqref{coadjoint J} (see \eqref{coadjoint P soft part} and \eqref{transfo soft part J}), they also transform in the coadjoint representation of $\mathfrak{bms}_{4}$ for the appropriate pairing \eqref{pairing hard soft}.

We define the bracket between the BMS fluxes as
\begin{equation}
    \{ F_{(\mathcal{T}_1, \mathcal{Y}_1, \bar{\mathcal{Y}}_1)} , F_{(\mathcal{T}_2, \mathcal{Y}_2, \bar{\mathcal{Y}}_2)}  \} = \delta_{(\mathcal{T}_1, \mathcal{Y}_1, \bar{\mathcal{Y}}_1)} F_{(\mathcal{T}_2, \mathcal{Y}_2, \bar{\mathcal{Y}}_2)} .
    \label{dirac bracket}
\end{equation} As shown in \cite{Compere:2020lrt}, the BMS fluxes \eqref{BMS flux} satisfy the algebra
\begin{equation}
   \{ F_{(\mathcal{T}_1, \mathcal{Y}_1, \bar{\mathcal{Y}}_1)} , F_{(\mathcal{T}_2, \mathcal{Y}_2, \bar{\mathcal{Y}}_2)} \} = - F_{[(\mathcal{T}_1, \mathcal{Y}_1, \bar{\mathcal{Y}}_1), (\mathcal{T}_2, \mathcal{Y}_2, \bar{\mathcal{Y}}_2)]},
   \label{current algebra}
\end{equation} which implies that the bracket \eqref{dirac bracket} corresponds to the Kirillov-Kostant Poisson bracket on $\mathfrak{bms}^*_4$. In terms of the momentum fluxes, the bracket \eqref{dirac bracket} can be written explicitly as
\begin{equation}
\begin{split}
    \{\bar{\mathcal{J}}(z, \bar{z}),\bar{\mathcal{J}}(w, \bar{w})\} &=  \delta^2 (z- w) {\partial}_w \bar{\mathcal{J}}(w, \bar{w}) + 2{\partial}_w \delta^2 (z-w) \bar{\mathcal{J}}(w, \bar{w}), \\
    \{\bar{\mathcal{J}}(z, \bar{z}),{\mathcal{J}}(w, \bar{w})\} &= \delta^2 (z- w) {\partial}_w \bar{\mathcal{J}}(w, \bar{w}) + {\partial}_w \delta^2 (z-w) \bar{\mathcal{J}}(w, \bar{w}), \\
    \{\mathcal{P}(z, \bar{z}), \bar{\mathcal{J}}(w,\bar{w})\} &= \frac{1}{2} \delta^2 (z-w) \partial_w \mathcal{P}(w, \bar{w}) + \frac{3}{2} \partial_w \delta^2 (z-w) \mathcal{P}(w,\bar{w}), \\
    \{\bar{\mathcal{J}}(z, \bar{z}), \mathcal{P}(w, \bar{w})\} &=  \delta^2 (z- w) \partial_w \mathcal{P}(w, \bar{w}) + \frac{3}{2} \partial_w \delta^2 (z-w) {\mathcal{P}}(w, \bar{w}), \\
    \{\mathcal{P}(z, \bar{z}), \mathcal{P}(w,\bar{w})\} &= 0,
    \end{split}
    \label{OPE fluxes classical}
\end{equation} together with the complex conjugate relations. In particular, these relations reproduce the desired variations \eqref{coadjoint P} and \eqref{coadjoint J}.  Similarly, the flux algebra \eqref{current algebra} can be written in terms of the momentum fluxes as
\begin{equation}
   \begin{split}
    &\{ \bar{\mathcal{J}}(z, \bar{z}), \bar{\mathcal{J}}(w, \bar{w} ) \} = - [\bar{\mathcal{J}}(z,\bar{z}) + \bar{\mathcal{J}}(w, \bar{w}) ] \partial_z \delta^2 (z-w),  \\
    &\{ \bar{\mathcal{J}}(z, \bar{z}), {\mathcal{J}}(w, \bar{w} ) \} = -[ \partial_z \delta^2 (z-w) \mathcal{J}(z, \bar{z}) - \partial_{\bar{z}}\delta^2 (z-w) \bar{\mathcal{J}}(w, \bar{w}) ], \\
    &\{ \mathcal{P}(z, \bar{z}),\bar{{\mathcal{J}}}(w, \bar{w}) \} = -[\frac{1}{2} \mathcal{P}(z, \bar{z}) + \mathcal{P}(w, \bar{w})] \partial_z  \delta^2 (z-w),  \\
    &\{ \mathcal{P}(z, \bar{z}),\mathcal{P}(w, \bar{w}) \} = 0 ,
   \end{split}
   \label{structure constant algebra}
\end{equation} together with the complex conjugate relations.

The bracket that has been considered up to this point is associated with the total BMS flux \eqref{BMS flux}. However, as discussed around equation \eqref{pairing hard soft} above, one could study the soft/hard sectors separately and consider the appropriate induced bracket on each of them. More explicitly, assuming that the soft and hard sectors factorize \cite{Campiglia:2021bap}, we have
\begin{equation}
    \{ F^{soft}_{(\mathcal{T}_1, \mathcal{Y}_1, \bar{\mathcal{Y}}_1)} , F^{hard}_{(\mathcal{T}_2, \mathcal{Y}_2, \bar{\mathcal{Y}}_2)}\} = 0, \qquad \{ F^{soft/hard} _{(\mathcal{T}_1 , \mathcal{Y}_1, \bar{\mathcal{Y}}_1)} , F_{(\mathcal{T}_2, \mathcal{Y}_2, \bar{\mathcal{Y}}_2)}^{soft/hard}   \}  = - F_{[(\mathcal{T}_1, \mathcal{Y}_1, \bar{\mathcal{Y}}_1), (\mathcal{T}_2, \mathcal{Y}_2, \bar{\mathcal{Y}}_2)]}^{soft/hard},
    \label{factorization hard soft}
\end{equation} where the second relation is straightforwardly obtained by using the first of \eqref{factorization hard soft}, the definition of the bracket \eqref{dirac bracket} and the results \eqref{coadjoint P soft part} and \eqref{transfo soft part J}. Henceforth, \eqref{OPE fluxes classical} and \eqref{structure constant algebra} can be specified to soft/hard sectors separately. In the following, since we want to relate the BMS flux algebra with the CCFT currents, we will focus on the soft sector. 

\section{Momentum fluxes and CCFT operators}
\label{Relation with supertranslation operator and stress-tensor}

One of the starting points of celestial holography was the remarkable observation that Weinberg's leading soft graviton theorem could be reformulated as the Ward identity arising from the insertion of a $(\frac{3}{2}, \frac{1}{2})$ Kac-Moody current $P(z,\bar{z})$, called the supertranslation operator \cite{Strominger:2013jfa,He:2014laa}. Similarly, it was later shown that the subleading soft graviton theorem \cite{Cachazo:2014fwa} could be rewritten as an insertion of a $(2,0)$ operator $T(z)$, identified as the stress-tensor of the celestial CFT, reproducing the Ward identity of a $2d$ CFT \cite{Kapec:2016jld}. Although the supertranslation operator $P(z,\bar{z})$ and the stress-tensor $T(z)$ play a fundamental role in celestial holography, their precise relation to the bulk solution space in presence of superrotations (with the inclusion of $N_{AB}^{vac}$ for consistency of the phase space) and their transformation properties under the extended BMS symmetries have not yet been explicitly worked out. In this Section, we explore these aspects and relate the BMS momentum fluxes introduced above with these CCFT operators.

The supertranslation operator $P(z,\bar{z})$ and its complex conjugate $\bar{P}(z,\bar{z})$ of weights $(\frac{3}{2}, \frac{1}{2})$ and $(\frac{1}{2}, \frac{3}{2})$, respectively, can be related to the soft supermomentum flux $\mathcal{P}_{soft}(z, \bar{z})$ as 
\begin{equation}
    \mathcal{P}_{soft}(z, \bar{z}) = \bar{\mathscr{D}} P(z, \bar{z})  + \mathscr{D} \bar{P}(z, \bar{z}) ,
    \label{Supertranslation operator}
\end{equation} where the derivative operators $\mathscr{D}$ and $\bar{\mathscr{D}}$ are defined in \eqref{derivative operators conformal}. From \eqref{soft hard P} and \eqref{Supertranslation operator}, one deduces the explicit expression of $P(z, \bar{z})$ and $\bar{P}(z, \bar{z})$ in terms of the bulk metric:
\begin{equation}
\begin{split}
    P(z, \bar{z}) &=  \bar{\mathscr{D}} \mathscr{N}^{(0)} \\
    &=  \frac{1}{16 \pi G} \int^{+\infty}_{-\infty}  du \, (\Omega_S\bar{\Omega}_S)^{\frac{1}{2}}   \Big[\Big({D}_{\bar{z}}+ \frac{1}{2} \bar{\partial}{\Phi} \Big)\hat{N}_{{{z}}{{z}}} \Big] \\
    &= \frac{1}{16 \pi G} (\Omega_S\bar{\Omega}_S)^{\frac{1}{2}} \Big({D}_{\bar{z}}+ \frac{1}{2} \bar{\partial}{\Phi} \Big) [ \Delta C N_{{z}{z}}^{vac} - 2 D^2_{{z}} \Delta C ] ,
    \end{split}
    \label{expression supertranslation operator soluton}
\end{equation} together with the complex conjugate expression for $\bar{P}(z, \bar{z})$. To obtain the last equality, we used the falloffs \eqref{falloff in u}. Comparing with \eqref{integrated expression P}, one can rewrite \eqref{Supertranslation operator} as 
\begin{equation}
     \mathcal{P}_{soft}(z, \bar{z}) = 2 \bar{\mathscr{D}} P(z, \bar{z})  = 2 \mathscr{D} \bar{P}(z, \bar{z}),
\end{equation} which is a direct consequence of the electricity condition \eqref{electricity condition}. Notice that the expression of the supertranslation operator \eqref{expression supertranslation operator soluton} that we are using is compatible with the one initially proposed in \cite{Strominger:2013jfa, He:2014laa} when setting $N_{AB}^{vac} = 0$. The additional terms that we have allow us to have nicer transformation laws under the action of BMS symmetries. In particular, the supertranslation operator is an actual Virasoro primary rather then a descendent thanks to the use of the derivative operators \eqref{derivative operators conformal}. Indeed, an explicit computation gives
\begin{equation}
        \{ F^{soft}_{(\mathcal{T}, \mathcal{Y}, \bar{\mathcal{Y}})} , P\} =  \delta_{(\mathcal{T}, \mathcal{Y}, \bar{\mathcal{Y}})} P = \mathcal{Y} \partial P + \bar{\mathcal{Y}} \bar{\partial} P + \frac{3}{2} \partial  \mathcal{Y} P  + \frac{1}{2}  \bar{\partial} \bar{\mathcal{Y}} P
\end{equation} or, equivalently,
\begin{equation}
    \begin{split}
        \{ \mathcal{P}_{soft}(z,\bar{z}), P(w, \bar{w}) \} &= 0, \\
        \{ \bar{\mathcal{J}}_{soft}(z,\bar{z}), P(w, \bar{w}) \} &= \delta^2(z-w) \partial_w P(w, \bar{w}) + \frac{3}{2} \partial_w \delta^2 (z-w) P(w, \bar{w}), \\
         \{ {\mathcal{J}}_{soft}(z,\bar{z}), P(w, \bar{w}) \} &= \delta^2(z-w) \partial_{\bar{w}} P(w, \bar{w}) + \frac{1}{2} \partial_{\bar{w}} \delta^2 (z-w) P(w, \bar{w}).
         \label{commutations for P operator}
    \end{split}
\end{equation} We have the analogous results for $\bar{P}$.

The stress-tensor of the CCFT is encoded in the complex conformal fields $T(z)$ and $\bar{T}(\bar{z})$ of weights $(2,0)$ and $(0,2)$, respectively. We observe that it can be constructed from the super angular momentum flux introduced in Section \ref{Generators and momenta} as 
\begin{equation}
    T(z) = \oint_{\mathcal{C}}\frac{d\bar{z}}{2i\pi} \, \bar{\mathcal{J}}_{soft}(z, \bar{z}), \qquad \bar{T}(\bz)= \oint_{\mathcal{C}} \frac{d z}{2i\pi} \, \mathcal{J}_{soft}(z, \bar{z}).
    \label{def stress tensor}
\end{equation} 
%These integrals correspond to the light-ray transforms introduced in Equation \eqref{light-ray transform}. 
Since total derivatives can be dropped out, the definition \eqref{def stress tensor} does not depend on the particular representatives of $\mathcal{J}_{soft}(z, \bar{z})$ and $\bar{\mathcal{J}}_{soft}(z, \bar{z})$ in the equivalent classes \eqref{equivalence classes}. The stress-tensor is related to the soft part of the flux for superrotations in \eqref{pairing hard soft} through
\begin{equation}
    F_{\mathcal{Y}}^{soft} =  \int_{\mathcal{S}} \frac{dz d\bar{z}}{(2i\pi)^2} \, \mathcal{Y} (z) \bar{\mathcal{J}}_{soft}(z, \bar{z}) = \oint_{\mathcal{C}} \frac{dz}{2i\pi} \, \mathcal{Y} (z) T(z)
    \label{soft charge in terms of T}
\end{equation} and the complex conjugate relation for $F_{\bar{\mathcal{Y}}}^{soft}$. One can recognize the last expression in \eqref{soft charge in terms of T} as the soft part of the superrotation charge \cite{Kapec:2014opa,Kapec:2016jld}. From \eqref{hard soft J} and \eqref{def stress tensor}, one deduces the explicit expression of $T$ in terms of the bulk metric:
\begin{equation}
\begin{split}
    T(z) &=  - \frac{1}{16\pi G} \oint_{\mathcal{C}} \frac{d\bar{z}}{2i\pi} \int^{+\infty}_{-\infty} du \, (\Omega_S \bar{\Omega}_S)^{-1} \Big[ u \Big( D_{{z}}^3 - 2 N_{{z}{z}}^{vac} {D}_{{z}} -  D_{{z}} N^{vac}_{{z}{z}}  \Big) \hat{N}^{{z}{z}}  \\ &\qquad\qquad\qquad\qquad+\hat{N}^{{z}{z}} D_{{z}}\Big[\Big(D^2_{{z}} - \frac{1}{2}  N_{{z}{z}}^{vac}  \Big) C_-\Big]+ 3 D_{{z}} \hat{N}^{{z}{z}} \Big( D_{{z}}^2 - \frac{1}{2} N_{{z}{z}}^{vac}  \Big) C_- \Big].
    \label{stress tensor in terms of the solution}
    \end{split}
\end{equation}  We have the complex conjugate expression for the anti-holomorphic $\bar{T}(\bar{z})$. The relation \eqref{stress tensor in terms of the solution} agrees with the one first proposed in \cite{Kapec:2016jld} when setting $N_{AB}^{vac} = 0 = C_-$. The decoration with the terms involving the memory fields is turned on once we are considering a vacuum that is not global Minkowski space \cite{Compere:2016jwb , Compere:2016hzt , Compere:2016gwf} and allows us to have nicer transformation laws. An explicit computation shows that
\begin{equation}
        \{ F^{soft}_{(\mathcal{T}, \mathcal{Y}, \bar{\mathcal{Y}})} , T\} =  \delta_{(\mathcal{T}, \mathcal{Y}, \bar{\mathcal{Y}})} T = \mathcal{Y} \partial T + 2 \partial  \mathcal{Y} T + \oint_{\mathcal{C}}\frac{d\bar{z}}{2i\pi} \, \Big[ \frac{1}{2} \mathcal{T} {\partial} \mathcal{P}_{soft} + \frac{3}{2} {\partial} \mathcal{T} \mathcal{P}_{soft} \Big]
\end{equation} or, equivalently,
\begin{equation}
    \begin{split}
        \{ \mathcal{P}_{soft}(z,\bar{z}), T(w) \} &= \frac{1}{2}\delta (z-w) \partial_w \mathcal{P}_{soft}(w, \bar{z}) + \frac{3}{2} \partial_w \delta (z-w) \mathcal{P}_{soft}(w, \bar{z}), \\
        \{ \bar{\mathcal{J}}_{soft}(z,\bar{z}), T(w) \} &= \delta^2(z-w) \partial_w T(w) + 2 \partial_w \delta^2 (z-w) T(w), \\
         \{ {\mathcal{J}}_{soft}(z,\bar{z}), T(w) \} &= 0.
    \end{split}
    \label{commutations for T operator}
\end{equation} We again have the analogous results for $\bar{T}$. 

Finally, using once more the definitions \eqref{Supertranslation operator} and \eqref{def stress tensor}, the commutation relations \eqref{commutations for P operator} and \eqref{commutations for T operator} lead to 
\begin{equation}
    \begin{split}
        \{P(z, \bar{z}), P(w, \bar{w})  \} &= 0, \\
        \{ P(z, \bar{z}) , \bar{P}(w, \bar{w})  \} &= 0, \\
        \{ T(z), P(w, \bar{w}) \} &= \delta (z-w) \partial_w P(w, \bar{w}) + \frac{3}{2} \partial_w \delta (z-w) P(w, \bar{w}), \\
        \{ \bar{T}(\bar{z}) ,  P(w, \bar{w}) \} &=  \delta (\bar{z}-\bar{w}) \partial_{\bar{w}} P(w, \bar{w}) + \frac{1}{2} \partial_{\bar{w}} \delta (\bar{z}-\bar{w}) P(w, \bar{w}), \\
        \{ T(z), T( w)  \} &= \delta (z-w) \partial_w T(w) + 2 \partial_w \delta (z-w) T(w), \\
        \{ \bar{T}(\bar{z}), T( w)  \} &=0. 
        \label{commutation relations CCFT operas}
    \end{split}
\end{equation} The other relations can be obtained by complex conjugation and antisymmetry of the bracket.

The conformal weights $(h, \bar{h})$ and spin/boost weights $(J, \Delta)$ of the relevant objects discussed in this paper are summarized in Table \ref{weights S} (see also \cite{Barnich:2021dta}). Under complex conjugation, we have
$\overline{(h, \bar{h})} = ( \bar{h} , h)$,
$\overline{(J,\Delta)} = (-J,\Delta)$.

\begin{table}[h]
\begin{center}
  \begin{tabular}{ | c | c | c | c | c | c | c | c | c | c | c | c |} \hline $\phi_{h,\bar{h}}$ & $
    {\mathcal T}$ & $ {\mathcal Y} $ & $ {\mathcal P} $ & $ {\mathcal J} $ & $\mathscr{N}^{(0)}$ & $\mathscr{N}^{(1)}$ & $\mathscr{C}$ &
    ${P}$ & $T$ & $\mathscr{D}$ & $dz$ \\ \hline $h$ & $-\frac{1}{2}$ & $-1 $ &
    $\frac{3}{2}$ & $1$ & $\frac{3}{2}$ & $1$ & $-\frac{1}{2}$ & $\frac{3}{2}$ & $2$ & $1$ & $-1$ \\ \hline $\bar h$ & $-\frac{1}{2}$ & $0$
    & $\frac{3}{2}$ & $2$ & $-\frac{1}{2}$ & $-1$ & $-\frac{1}{2}$ & $\frac{1}{2}$ & $0$ & $0$ & $0$  \\ \hline
    $J$ & $0$ & $-1$ & $0$ & $-1$ & $2$ & $2$ & $0$ & $1$ & $2$ & $1$ & $-1$ 
 \\ \hline
 $\Delta$ & $-1$ & $-1$ & $3$ & $3$ & $1$ & $0$ & $-1$ & $2$ & $2$ & $1$ & $-1$ \\ \hline \end{tabular} \caption{Conformal weights $(h, \bar{h})$ and spin/boost weights $(J, \Delta)$.} \label{weights S}
\end{center}
\end{table}

\section{Constraints on CCFT}

Up to this stage, all the results have been obtained from gravitational bulk computations. Some non-local combinations of the solution space have been identified as conformal fields on the celestial Riemann surface whose transformation laws are induced by bulk diffeomorphisms. In this construction, a phase space structure has emerged naturally from the BMS flux algebra. We now study the implications of these results at the quantum level and derive the OPEs between the various conformal operators.

Using standard arguments \cite{book:75861}, one can deduce the singular parts of the OPEs between the operators associated with BMS momentum fluxes by starting from their commutation relations \eqref{OPE fluxes classical}. We have explicitly
\begin{equation}
    \begin{split}
    \bar{\mathcal{J}}(z, \bar{z})\bar{\mathcal{J}}(w, \bar{w}) &\sim  \frac{2}{(z-w)^2 (\bar{z}-\bar{w})} \bar{\mathcal{J}}(w, \bar{w})+\frac{1}{(z-w)(\bar{z}-\bar{w})} {\partial}_w \bar{\mathcal{J}}(w, \bar{w}) ,\\
    \bar{\mathcal{J}}(z, \bar{z}){\mathcal{J}}(w, \bar{w}) &\sim  \frac{1}{(z-w)^2(\bar{z}-\bar{w})}  \bar{\mathcal{J}}(w, \bar{w})+\frac{1}{(z-w)(\bar{z}-\bar{w})} {\partial}_w \bar{\mathcal{J}}(w, \bar{w}) , \\
    \mathcal{P}(z, \bar{z}) \bar{\mathcal{J}}(w,\bar{w}) &\sim \frac{3/2}{(z-w)^2(\bar{z}-\bar{w})} \mathcal{P}(w, \bar{w})+  \frac{1/2}{(z-w)(\bar{z}-\bar{w})} \partial_w\mathcal{P}(w, \bar{w}), \\
    \bar{\mathcal{J}}(z, \bar{z}) \mathcal{P}(w, \bar{w}) &\sim \frac{3/2}{(z-w)^2(\bar{z}-\bar{w})}  \mathcal{P}(w, \bar{w})+ \frac{1}{(z-w)(\bar{z}-\bar{w})} \partial_w \mathcal{P}(w, \bar{w}), \\
    \mathcal{P}(z, \bar{z}) \mathcal{P}(w,\bar{w}) &\sim 0 ,
    \label{OPE fluxes}
    \end{split}
\end{equation}
together with the complex conjugate relations. The notation ``$\sim$'' means equality modulo expressions that are regular as $(z,\bar{z}) \to (w, \bar{w})$. The third OPE in \eqref{OPE fluxes} can be deduced from the fourth one using $\mathcal{P}(z, \bar{z}) \bar{\mathcal{J}}(w,\bar{w}) =  \bar{\mathcal{J}}(w,\bar{w}) \mathcal{P}(z, \bar{z})$. We will avoid writing redundant OPEs in the following.  

As a consequence of the factorization between hard and soft sectors in the phase space discussed at the end of Section \ref{sec:BMS flux algebra}, the above OPEs can be written for hard and soft BMS momentum fluxes separately. From now on, since we want to find constraints on the CCFT from celestial currents, we will restrict ourselves to the soft sector. The OPEs between soft BMS momentum fluxes and the CCFT operators can be readily deduced from \eqref{commutations for P operator} and \eqref{commutations for T operator}, leading to:
\begin{equation}
    \begin{split}
         \mathcal{P}_{soft}(z,\bar{z}) P(w, \bar{w}) &\sim 0, \\
         \bar{\mathcal{J}}_{soft}(z,\bar{z}) P(w, \bar{w}) &\sim \frac{3/2}{(z-w)^2(\bar{z}- \bar{w})}  P(w, \bar{w})+\frac{1}{(z-w)(\bar{z}- \bar{w})} \partial_w P(w, \bar{w}),   \\
         {\mathcal{J}}_{soft}(z,\bar{z}) P(w, \bar{w}) &\sim \frac{1/2}{(z-w)(\bar{z}- \bar{w})^2} P(w, \bar{w}) +\frac{1}{(z-w)(\bar{z}- \bar{w})} \partial_{\bar{w}} P(w, \bar{w}),  \\
         \mathcal{P}_{soft}(z,\bar{z}) T(w)  &\sim \frac{3/2}{(z-w)^2} \mathcal{P}_{soft}(w, \bar{z})+\frac{1/2}{(z-w)} \partial_w \mathcal{P}_{soft}(w, \bar{z}), \\
     \bar{\mathcal{J}}_{soft}(z,\bar{z}) T(w) &\sim \frac{2}{(z-w)^2(\bar{z}-\bar{w})}  T(w)+\frac{1}{(z-w)(\bar{z}-\bar{w})} \partial_w T(w), \\
         {\mathcal{J}}_{soft}(z,\bar{z}) T(w) &\sim 0.
         \label{OPE BMS CCFT}
    \end{split}
\end{equation} 
Finally, using the commutation relations \eqref{commutation relations CCFT operas}, one obtains the OPEs between CCFT operators:
\begin{equation}
    \begin{split}
        P(z, \bar{z}) P(w, \bar{w})  &\sim 0, \\
         P(z, \bar{z}) \bar{P}(w, \bar{w})  &\sim 0, \\
        T(z) P(w, \bar{w}) &\sim \frac{3/2}{(z-w)^2} P(w, \bar{w}) +\frac{1}{(z-w)} \partial_w P(w, \bar{w}),  \\
        \bar{T}(\bar{z})  P(w, \bar{w}) &\sim  \frac{1/2}{(\bar z-\bar w)^2} P(w, \bar{w})+\frac{1}{(\bar{z}-\bar w)} \partial_{\bar{w}} P(w, \bar{w}), \\
        T(z) T( w)&\sim  \frac{2}{(z-w)^2}T(w)+\frac{1}{(z-w)} \partial_w T(w), \\
        \bar{T}(\bar{z}) T( w) &\sim 0.
        \label{OPEs relations CCFT operas}
    \end{split}
\end{equation} These results are compatible with those given in \cite{Barnich:2017ubf}. They also match with the OPEs found in \cite{Fotopoulos:2019vac} that were derived from collinear and conformally soft limits of amplitudes, up to the fact that the supertranslation operator that we are considering here is a Virasoro primary rather than a descendant.

Let us now elaborate more on the constraints involving the celestial CFT operators and momentum flux operators with generic conformal operators. To simplify the discussion, we set the memory fields to zero, i.e. $\varphi = 0 = \bar{\varphi}$ ($N_{AB}^{vac} = 0$) and $C_- =0$. In celestial holography, a massless particle of energy $\omega$ involved in a scattering process in $4d$ flat space is associated to an operator $\mathcal O(\omega,z,\bz)$ (which can depend on other quantum numbers, which are omitted in this notation), where $(z,\bz)$ labels the point on the celestial sphere where the particle exits (or enters) spacetime \cite{He:2015zea,Strominger:2017zoo}. Instead of working in the usual momentum basis, a promising celestial dictionary  involves the Mellin representation \cite{Pasterski:2016qvg,Pasterski:2017ylz,Pasterski:2017kqt}
\begin{equation}\label{cele}
    \mathcal O_{h, \bar{h}}(z,\bz) =\int_0^\infty d\omega\, \omega^{\Delta-1} \mathcal O(\omega,z,\bz),
\end{equation}
which trades the energy $\omega$ for the conformal dimension $\Delta=h+\bar h$. Celestial operators \eqref{cele} indeed enjoy the property of transforming as $2d$ quasi-primaries. The so-called conformally soft limits \cite{Donnay:2018neh}, for which the conformal dimension takes specific values, lead to $2d$ currents on the CCFT; see e.g.
\cite{Donnay:2018neh,Nandan:2019jas,Himwich:2019dug,Pate:2019lpp,Donnay:2020guq,Banerjee:2020kaa,Pasterski:2021fjn,Pasterski:2021dqe,Guevara:2021abz,Strominger:2021lvk,Himwich:2021dau}.

It was shown that the OPEs involving the components of the CCFT stress-tensor\footnote{This operator is actually obtained from the shadow of the conformally soft operator with $\Delta=0$ \cite{Cheung:2016iub,Donnay:2018neh}.} $T(z)$, $\bar{T}(\bar{z})$ and celestial operators   $\mathcal O_{h, \bar{h}}$ representing gauge bosons and gravitons are given by
\begin{equation}
    \begin{split}
        T(z) \mathcal O_{h, \bar{h}}(w,\bar{w}) &\sim \frac{h}{(z-w)^2} \mathcal O_{h, \bar{h}}(w,\bar{w})+\frac{1}{(z-w)} \partial_w \mathcal O_{h, \bar h}(w,\bar{w}), \\
        \bar{T}(z) \mathcal O_{h, \bar{h}}(w,\bar{w}) &\sim \frac{\bar h}{(\bz-\bw)^2} \mathcal O_{h, \bar{h}}(w,\bar{w})+\frac{1}{(\bar{z}-\bar{w})} \partial_{\bar{w}} \mathcal O_{h, \bar h}(w,\bar{w}),
    \end{split}\label{OPE stress tensor phi}
\end{equation} 
which implies that these operators are Virasoro primaries.
These expressions were derived from collinear and conformally soft limits of Einstein-Yang-Mills amplitudes in \cite{Fotopoulos:2019tpe,Fotopoulos:2019vac}.
We deduce from \eqref{OPE stress tensor phi} that OPEs involving the super angular momentum flux are of the following form:
\begin{equation}
    \begin{split}
         \bar{\mathcal{J}}(z, \bar{z}) \mathcal O_{h, \bar{h}}(w,\bar{w}) &\sim \frac{h}{(z-w)^2(\bar{z}-\bar{w})} \mathcal O_{h, \bar{h}}(w,\bar{w})+ \frac{1}{(z-w)(\bar{z}-\bar{w})} \partial_w \mathcal O_{h, \bar h}(w,\bar{w}),
         \\
         {\mathcal{J}}(z, \bar{z}) \mathcal O_{h, \bar{h}}(w,\bar{w}) &\sim \frac{\bar h}{(z-w)(\bar{z}-\bar{w})^2} \mathcal O
         _{h, \bar{h}}(w,\bar{w})+ \frac{1}{(z-w)(\bar{z}-\bar{w})} \partial_\bw \mathcal O_{h, \bar h}(w,\bar{w}). 
    \end{split}
\end{equation} 

While superrotations lead to the expected expressions \eqref{OPE stress tensor phi} in a CFT, it is a notorious fact in celestial holography that supertranslation symmetry is more subtle to deal with. It particular, it has been shown that the insertion of the supertranslation operator $P(z, \bar{z})$ into a celestial correlation function gives \cite{Donnay:2018neh,Fotopoulos:2019vac}: 
\begin{equation}\label{OPEP}
\begin{split}
    P(z, \bar{z}) \mathcal O_{h, \bar{h}}(w, \bar{w}) &\sim \frac{1}{(z-w)} \mathcal O_{h+ \frac{1}{2}, \bar{h}+ \frac{1}{2}}(w, \bar{w}) .\\
    %\bar{P}(z, \bar{z}) \phi_{h, \bar{h}}(w, \bar{w}) &\sim \frac{1}{(\bar{z}-\bar{w})} \phi_{h+ \frac{1}{2}, \bar{h}+ \frac{1}{2}}(w, \bar{w}) .
\end{split}
\end{equation} 
This OPE relationship is nothing but the celestial consequence of Weinberg's leading soft graviton theorem, as it can be readily obtained from a Mellin transform of the Ward identity associated to supertranslation symmetry \cite{Strominger:2013jfa,He:2014laa}.
As one can see from \eqref{OPEP}, the action of supertranslations (even global ones) leads to a shift of $(\12,\12)$ in the conformal weights of celestial operators.
Now one can deduce\footnote{To do so, we act on \eqref{OPEP} with the formal anti-derivative $2\pi\partial_\bz^{-1}=\int d^2y \frac{1}{z-y}$ and make use of formulae for $2d$ conformal integrals as in \cite{Osborn:2012vt}.} the generic form of the OPE between the supermomentum flux operator $\mathcal{P}(z, \bar{z})$ and a celestial operator $\mathcal O_{h, \bar{h}}(w, \bar{w})$:
\begin{equation}
    \mathcal{P}(z, \bar{z}) \mathcal O_{h,\bar{h}} (w, \bar{w}) \sim \frac{1}{(z-w)(\bar{z} - \bar{w})} \mathcal O_{h+ \frac{1}{2}, \bar{h}+ \frac{1}{2}}(w, \bar{w}).
\end{equation} This formula agrees with the expression found in \cite{Fotopoulos:2019vac}, which was obtained by taking successive commutators involving the zero-mode of $P$ and the stress-tensor. 

\section{Discussion}

We now conclude by discussing some possible extensions and consequences of the results presented in this work.

\paragraph{Surface charges versus fluxes} We have argued that the fluxes are more natural objects from the point of view of the CCFT than the surface charges at fixed value of the retarded time $u$. In particular, the fluxes are completely determined by the values of the surface charges at the corners $\mathscr{I}^+_\pm$ of null infinity, which are non-radiative regions of the spacetime. The statement of the closure of the flux algebra \eqref{current algebra} can then be recast as the closure of the surface charge algebra at $\mathscr{I}^+_\pm$, without the need of modified bracket or the appearance of $2$-cocycle. This echoes with recent works suggesting that symmetries are encoded at the corners of hypersurfaces \cite{Donnelly:2016auv,Laddha:2020kvp,Freidel:2020xyx,Freidel:2020svx,Freidel:2020ayo,Donnelly:2020xgu,Ciambelli:2021vnn}. The price to pay for considering fluxes at $\mathscr{I}^+$ instead of surface charges is that we lose information on the local flux-balance laws such as the Bondi mass loss formula. Henceforth, both point of views are complementary: the integrated fluxes describe the state of a system, while the surface charges point of view provides information on its dynamics.

\paragraph{Central charge in the CCFT} In the AdS$_3$/CFT$_2$ correspondence, the central charge of the boundary CFT$_2$ can be read from the Brown-Henneaux central extension \cite{Brown:1986nw}. The latter appears classically by computing the charge algebra of large diffeomorphisms in asymptotically AdS$_3$ spacetimes. One might expect that a similar feature would hold in the present context, namely that the possible CCFT$_2$ central extension appears in the classical bulk computation of the charge algebra. However, as stated in \eqref{current algebra}, the BMS flux algebra closes under the standard Peierls bracket and does not exhibit a central term. This indicates that at  least the garden-variety type of central charge of the CCFT$_2$ vanishes, which is in agreement with the results found in \cite{Fotopoulos:2019vac} from computing the $T T$ OPE (see also \eqref{OPEs relations CCFT operas}). Let us notice that there is still an imprint of central extension in the transformation of the Liouville stress-tensor that exhibits a Schwarzian derivative (see equation \eqref{transfo of the 2d Liouville tens}). 

\paragraph{From extended to generalized BMS}

In this paper, we have considered gravity in asymptotically flat spacetimes with a fixed boundary structure. Allowing for some singular punctures on the celestial Riemann surface enables to include the whole $\text{Witt}\oplus\overline{\text{Witt}}$ superrotations in the asymptotic symmetry algebra, leading to the extended BMS algebra \cite{Barnich:2010eb,Barnich:2009se,Barnich:2011ct}. If fluctuations of the transverse boundary metric $q_{AB}dx^A dx^B$ are permitted on the phase space (but keeping a fixed determinant $\sqrt{q} = \sqrt{\mathring{q}}$), one gets instead the generalized BMS algebra Diff($S^2$)$\loplus\mathfrak{s}$ where the superrotations Diff($S^2$) are smooth diffeomorphisms on the celestial sphere \cite{Campiglia:2015yka,Campiglia:2014yka,Compere:2018ylh,Flanagan:2019vbl}. Note that these smooth superrotations can be extended to all diffeomorphisms with isolated singularities, written Diff$(S^2)^*$, if the celestial sphere admits some punctures. The latter singular extension is relevant when considering Ward identities of the S-matrix and the relation with subleading soft graviton theorems \cite{Kapec:2014opa,Campiglia:2015yka,Campiglia:2014yka,Donnay:2020guq}. In particular, we have $\text{Witt}\oplus\overline{\text{Witt}}\subset$ Diff$(S^2)^*$, which implies that a notion of conformal field can still be defined: a conformal field $\phi_{h, \bar{h}}(z,\bar{z})$ transforms as in \eqref{def conformal field} under the subgroup of conformal coordinate transformations accompanied by a compensating Weyl rescaling to maintain a fixed determinant. The fluxes of charges associated with generalized BMS take the form
\begin{equation}
    F^{gen}_{(\mathcal{T}, \mathcal{Y}, \bar{\mathcal{Y}})} =  \int_{\mathcal{S}} \frac{dzd\bar{z}}{(2i\pi)^2} \, [\mathcal{T} \mathcal{P} + \mathcal{Y} \bar{\mathcal{J}} + \bar{\mathcal{Y}} \mathcal{J} ],
    \label{pairing gen bms}
\end{equation} where the superrotation generators $\mathcal{Y}= \mathcal{Y}(z, \bar{z})$ are now (possibly singular) functions of $(z, \bar{z})$ on the celestial Riemann surface. Consequently, since there is no constrain on $\mathcal{Y}$,  one does not need to quotient the super angular momentum fluxes $\mathcal{J}(z, \bar{z})$ and $\bar{\mathcal{J}}(z, \bar{z})$ by the equivalence relation \eqref{equivalence classes}. In this generalized BMS case, the precise expressions of $\mathcal{P}$, $\mathcal{J}$ and $\bar{\mathcal{J}}$ in terms of the solution space and their split into hard/soft sectors can be deduced from the analysis displayed in e.g. \cite{Compere:2018ylh,Campiglia:2020qvc,Compere:2020lrt}. They should be such that \eqref{coadjoint P}, \eqref{coadjoint P soft part}, \eqref{coadjoint J} and \eqref{transfo soft part J} still hold, so that the results of Section \ref{sec:BMS flux algebra} concerning the BMS flux algebra remain valid. 

As argued in \cite{Donnay:2020guq}, a natural object in this context is the shadow transform\footnote{The shadow transform is defined as $\widetilde{\mathcal O}_{1-h,1-\bar h}(w,\bw)=\int d^2z \, (w-z)^{2h-2}(\bw-\bz)^{2\bar h-2}\mathcal O_{h,\bar h}(z,\bz)$ \cite{Osborn:2012vt}, up to a normalization factor which can be chosen in such a way the shadow operation squares to one.} of the stress-tensor, which leads to an operator $\widetilde T$ of weights $(-1,1)$:
\begin{equation}\label{Ttilde}
    \widetilde{T}(w, \bar{w}) = \frac{3}{2\pi}\int d^2z  \frac{(w-z)^2}{(\bar{w} - \bar{z})^2} T(z).
\end{equation} 
It has the following OPE\footnote{It can be obtained from shadowing the $T(z)$ OPE, or read from expressions involving the $\Delta \to 0$ conformally soft operator in Refs. \cite{Cheung:2016iub,Adamo:2019ipt,Fotopoulos:2019vac}.}:
\begin{equation}
\widetilde{T}(w, \bar{w}) \mathcal O_{h,\bar h}(z,\bz) \sim \frac{-2h (w-z)}{(\bw-\bz)}\mathcal O_{h,\bar h}(z,\bz)+\frac{ (w-z)^2}{(\bw-\bz)}\p_z \mathcal O_{h,\bar h}(z,\bz).
\end{equation}
Using the observations relating the shadow stress-tensor with the soft part of the superrotation charge that were made in \cite{Cheung:2016iub,Donnay:2020guq}, \eqref{Ttilde} can be identified (up to a factor) with  $F^{gen,soft}_{\mathcal{Y}=\frac{(z-w)^2}{(\bz-\bw)}}$.  
It would be interesting to explore the shadow supermomenta and super angular momenta, which are operators of weights  $(-\12,-\12)$ and $(0,-1)$. More generally, it would be enlightening to have a full control of shadow transformations from the bulk gravitational phase space point of view.

\paragraph{From generalized to Weyl BMS} A new enhancement of the generalized BMS symmetries has been found recently by allowing the determinant $\sqrt{q}$ of the boundary metric to fluctuate on the phase space \cite{Freidel:2021fxf}. This asymptotic symmetry algebra contains the Weyl rescaling symmetries discussed in \cite{Barnich:2010eb, Barnich:2016lyg , Barnich:2019vzx} and has the following semi-direct structure:
\begin{equation}
    [\text{Diff($S^2$)}\loplus \text{Weyl}  ] \loplus \mathfrak{s}.
\end{equation} It was shown that the charges and fluxes associated with Weyl rescaling symmetries were non-vanishing, which suggests the presence of an additional term in the flux \eqref{pairing gen bms}. It would be interesting to repeat the analysis of the present paper for this case and use the framework developed in \cite{Barnich:2021dta,Donnelly:2020xgu} to treat conformal coordinate transformations and Weyl rescalings separately. 

\section*{Acknowledgments}
We are grateful to Glenn Barnich, Geoffrey Compère, Adrien Fiorucci, Laurent Freidel, Gaston Giribet, Prahar Mitra and Andy Strominger for useful discussions and comments. L.D. is supported by the
European Union's Horizon 2020 research and innovation programme under the Marie Sk{\l}odowska-Curie grant agreement No. 746297 and by the Austrian Science Fund (FWF), project P 30822. R.R. is supported by the
FWF, project P 32581-N.

\bibliographystyle{style}
\bibliography{references}

\end{document}